\documentclass[pre,showpacs,showkeys,preprintnumbers,amsmath,amssymb,superscriptaddress,twocolumn]{revtex4}
\usepackage{graphicx}
\usepackage{amsfonts}
\usepackage{url}
\usepackage{epstopdf}
\usepackage{color}
\usepackage{tikz-cd}
\usepackage{comment}
\usepackage[colorlinks=true,linkcolor=blue,citecolor=red]{hyperref}%
\usepackage{bm}

\def\R{{\mathbb R}}

\def\C{{\mathcal C}}

\def\c{c}

\def\x{\mathbf{x}}
\def\r{\bar{r}}

\def\ctanh{\mathrm{ctanh}}

\begin{document}

\title{Diffusion towards a nanoforest of absorbing pillars}

\author{Denis~S.~Grebenkov}
 \email{denis.grebenkov@polytechnique.edu}
\affiliation{
Laboratoire de Physique de la Mati\`{e}re Condens\'{e}e (UMR 7643), \\ 
CNRS -- Ecole Polytechnique, IP Paris, 91128 Palaiseau, France}

\author{Alexei T. Skvortsov}
\affiliation{
Maritime Division, Defence Science and Technology Group, 506 Lorimer Street, Fishermans Bend, Victoria 3207, Australia}

\date{\today}

\begin{abstract}
Spiky coatings (also known as nanoforests or Fakir-like surfaces) have
found many applications in chemical physics, material sciences and
biotechnology, such as superhydrophobic materials, filtration and
sensing systems, selective protein separation, to name but a few.  In
this paper, we provide a systematic study of steady-state diffusion
towards a periodic array of absorbing cylindrical pillars protruding
from a flat base.  We approximate a periodic cell of this system by a
circular tube containing a single pillar, derive an exact solution of
the underlying Laplace equation, and deduce a simple yet exact
representation for the total flux of particles onto the pillar.  The
dependence of this flux on the geometric parameters of the model is
thoroughly analyzed.  In particular, we investigate several asymptotic
regimes such as a thin pillar limit, a disk-like pillar, and an
infinitely long pillar.  Our study sheds a light onto the trapping
efficiency of spiky coatings and reveals the roles of pillar
anisotropy and diffusional screening.  
\end{abstract}





\maketitle

\section{Introduction}

\begin{figure}[h!]
\begin{center}
\includegraphics[width=80mm]{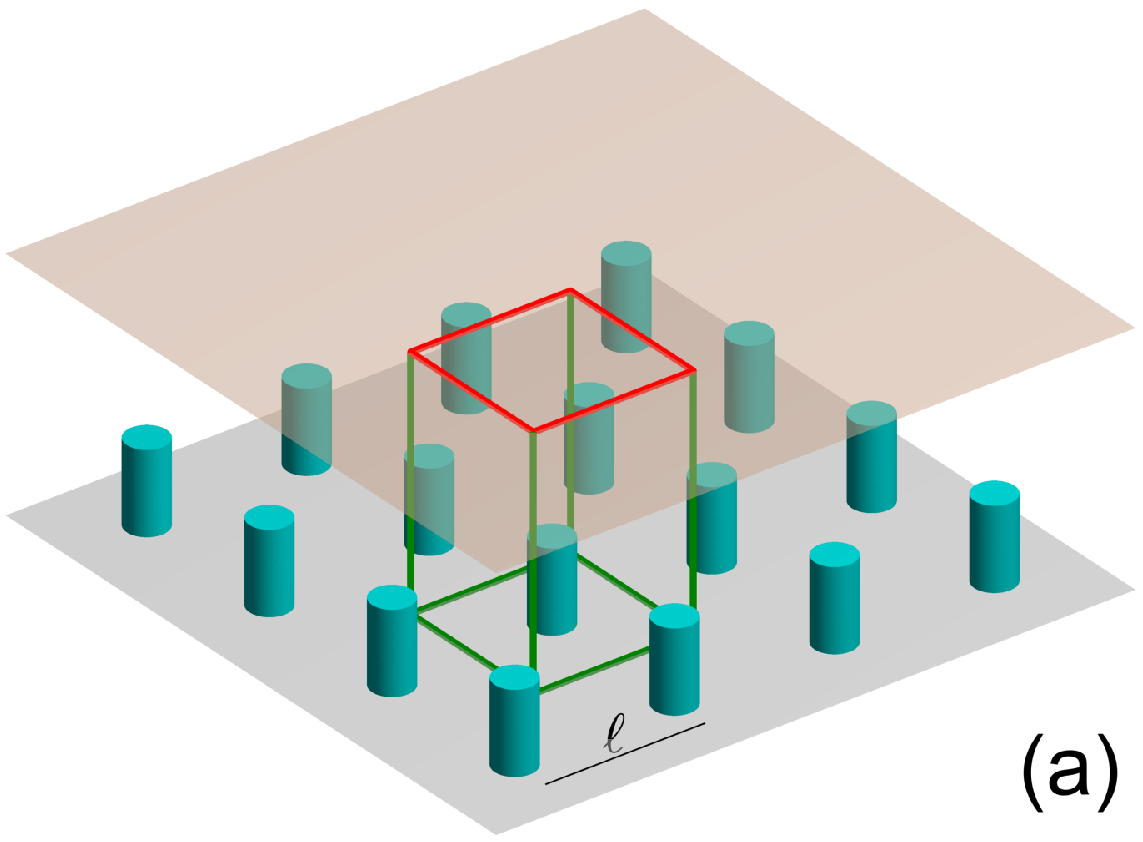} 
\includegraphics[width=35mm]{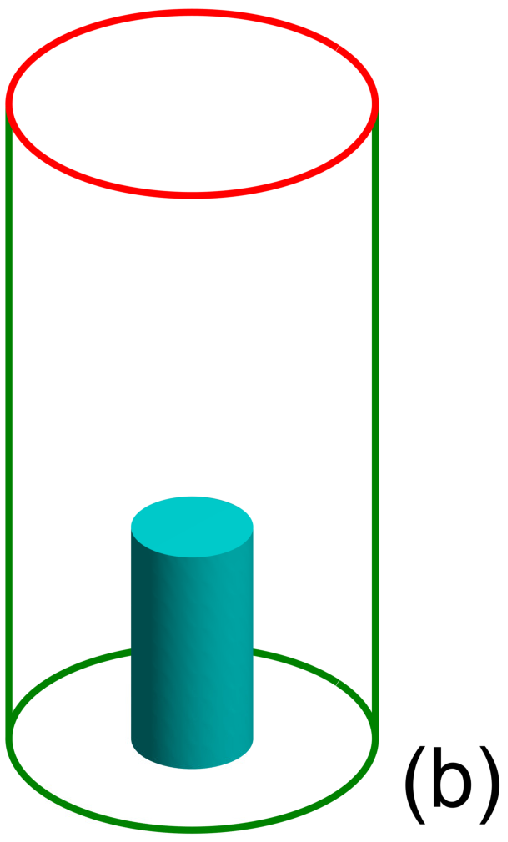} \hskip 5mm 
\includegraphics[width=30mm]{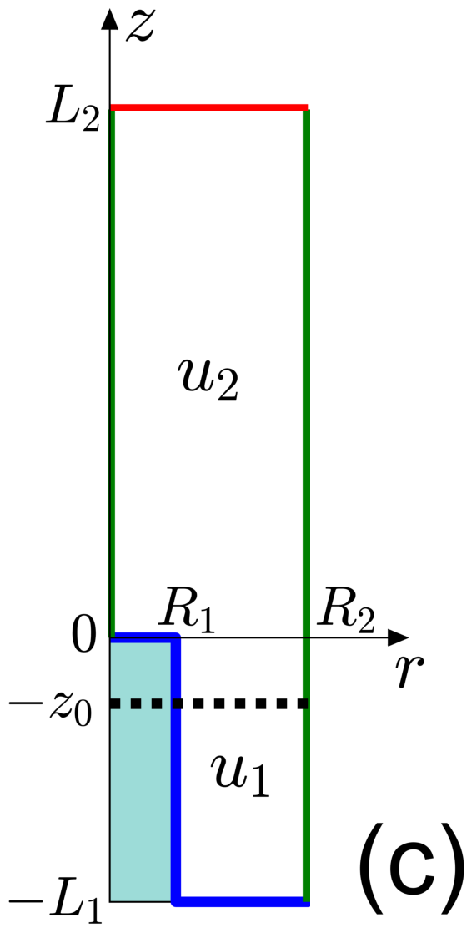} 
\end{center}
\caption{
{\bf (a)} A square-lattice array of cylindrical pillars (in light
blue) on a support (in gray) and an upper plane (in pink) that is a
source of particles.  Periodicity of this domain allows one to treat
diffusion in a single periodic cell around one pillar (shown by a
rectangular parallelepiped).  {\bf (b)} A single pillar surrounded by
an effective co-axial reflecting tube and capped by two parallel
planes; {\bf (c)} Planar ($xz$) projection of the three-dimensional
domain.  Red segment represents the source at the top, blue segments
show the absorbing pillar and partially reactive base, and green
segments indicate reflecting parts.  Shadowed (light blue) region is
the solid (inaccessible) interior of the cylindrical pillar.  Black
dotted line indicates the location $z = - z_0$ of the effective flat
absorbing surface that receives the same total flux as the spiky
coating (shown by blue segments).  Here $R_2$ is the radius of the
outer reflecting tube, $R_1$ is the radius of the absorbing pillar,
$L_1$ is its height, and $L_2$ is the distance between the source and
the top of the pillar (i.e., $L_1+L_2$ is the height of the whole
system).  Note that $R_2$ is related to the inter-pillar distance
$\ell$, e.g., $R_2 = \ell/\sqrt{\pi}$ for the square lattice.
Restrictions of the solution to the lower subdomain (with $-L_1 < z <
0$) and to the upper subdomain (with $0 < z < L_2$) are denoted by
$u_1$ and $u_2$, respectively. }
\label{fig:scheme}
\end{figure}

Diffusive transport towards irregular surfaces plays an important role
in various transport phenomena in nature and industry
\cite{Rice,Hughes,Redner_2001,Krapivsky,Schuss,Metzler14,Lindenberg19,Grebenkov07,Benichou14,Grebenkov_2020b},
including chemical engineering (inhomogeneous catalysis
\cite{benAvraham_2010,Coppens99,Filoche05,Filoche08b}, crystal growth
\cite{Witten81,Halsey90,Filoche00b,Bernard_2001}, colloidal physics
\cite{Goldstein91,Duplantier91}), geophysics (sedimentation
\cite{Garde_1977}, pollutant transport \cite{Boffetta_2008}, heat
conduction \cite{Blyth_1993,Blyth_2003,Andrade04,Rozanova12}),
electrochemistry
\cite{Liu85,Sapoval88,Blender90,Chassaing90,Halsey92,Grebenkov06}, 
microbiology (permeation through membranes or porous channels
\cite{Sapoval94,Levitz06}, intracellular transport and cell
signaling \cite{Lauffenburger_1993,Bressloff13,Hofling13}), and
physiology (gas exchange in lungs and placentas
\cite{Grebenkov_2005,Serov16,Sapoval21}).  There is a vast body of
literature on this subject, including the pioneering works of Berg,
Purcell, Zwanzig, Sapoval, and many others (see
\cite{Berg77,Zwanzig90,Zwanzig91,Sapoval94,benAvraham_2010} and
references therein).  Spiky interfaces (Fig. \ref{fig:scheme}), which
are formed by needles or pillars protruding from a base and often
referred to as nanoforests \cite{Kharisov_2015} or Fakir-like surfaces
\cite{Davis_2010}, have recently drown significant attention in many
areas of the so-called Laplacian transport due to the rapid progress
in fabrication technology and the favorable performance of spiky
coatings in many engineering applications such as superhydrophobic
materials \cite{Davis_2010}, filtration \cite{Ramon_2012,Ramon_2013},
sensing systems \cite{Nair_2007,Wei_2017,Chen_2022}, selective protein
separation \cite{Borberg19}, to name but a few.  In spiky coatings,
the diffusive flux has a local singularity near the tip of each spike
(resulting from a singular solution of the underlying Laplace
equation) and this amounts to a competition between the tips of the
thin spikes and the base of the coating for capturing the diffusing
particles.  For instance, in the two-dimensional setting with a dense
configuration of spikes (riblets), all flux is absorbed near the tips
of the spikes and the role of the coating base becomes negligible
\cite{Skvortsov_2014,Skvortsov_2018,Skvortsov_2018a}.  More
generally, the active zone, at which most of the Laplacian transport
takes place, has the fractal dimension equal to $1$ and thus scales
linearly with the size of the system, regardless its geometric
complexity \cite{Gutfraind93,Sapoval99,Filoche99,Filoche00}.  From the
mathematical point of view, this is a consequence of the Makarov
theorem on conformal mappings in the plane \cite{Makarov85,Jones88}.
As this fundamental result is known to fail in three dimensions
\cite{Bourgain87,Grebenkov05b}, the Laplacian transport towards spiky
coatings remains much less understood.  Actually, we are unaware of
any similar analytical results for the three-dimensional geometry,
which is the most relevant for applications, and this was one of the
motivations for the present study.

Mathematically the problem reduces to finding a solution to the
Laplace equation in a domain with geometrically complex boundaries.
The irregularity of the spiky profile makes the application of
conventional perturbation methods difficult.  For the two-dimensional
profiles (comb-like structures) this difficulty can be overcome by
means of conformal mapping
\cite{Bernard_2001,Vandembroucq_1997,Skvortsov_2014,Grebenkov16,Hewett_2016,Skvortsov_2018},
which is inapplicable in three dimensions.
One of the common methods for finding an {\it approximate} solution of
such type of problems in any dimension consists in replacing the
original irregular interface with an equivalent flat interface, which
provides the same diffusive flux through the system but is amenable to
analytical treatment
\cite{Sapoval94,Sarkar_1995,Vandembroucq_1997,Grebenkov06,Skvortsov_2014}.
Defining the equivalent interface for a particular geometry is not
straightforward and is actually the main challenge of this approach
(which is a variant of the effective medium approximation).  For
instance, Sarkar and Prosperetti \cite{Sarkar_1995} found an
equivalent interface for a flat surface covered by a sparse
configuration of hemispheroidal ``bosses''.  We emphasize that their
method relies on regularity of the interface profile that prohibits
its translation to the ``spiky'' limit (see Sec. \ref{sec:absorbing}).
Another example of an array of absorbing disks was discussed in Refs.
\cite{Bruna15,Lindsay_2017,Bernoff18,Bernoff18b,Paquin22}.

In this paper, we revisit the problem of steady-state diffusion
towards a spiky coating, i.e., an array of absorbing spikes protruding
from a flat base.  We model this structure as an infinite square
lattice of identical absorbing cylindrical pillars of radius $R_1$ and
height $L_1$, in which the centers of the closest pillars are
separated by the inter-pillar distance $\ell$
(Fig. \ref{fig:scheme}(a)).  We impose a constant concentration $\c_0$
of particles at a distance $L_2$ above the pillars and aim at finding
the total flux of particles onto each pillar, $J$, as a function of
four geometric parameters of this model: $L_1$, $R_1$, $L_2$ and
$\ell$.  For this purpose, we follow the rationale by Keller and Stein
\cite{Keller_1967} to reduce the original problem to that of diffusion
inside a cylindrical tube towards a single co-axial pillar
(Fig. \ref{fig:scheme}(b)).  The radius $R_2$ of the reflecting
boundary of that tube can be directly related to the inter-pillar
distance $\ell$ (see details below).  After that, we derive the exact
solution of the reduced problem and deduce the total flux as
\begin{equation}  \label{eq:Jexact}
J = \frac{\pi R_2^2 D \c_0}{L_2 + z_0} \,,
\end{equation}
where $D$ is the diffusion coefficient, and $z_0$ is the offset
parameter, which has the units of length and represents an aggregated
effect of the inhomogeneous coating, incorporating the entire coating
complexity (geometrical dimensions and non-uniform absorbing
properties).  This quantity, which depends only on the geometric
parameters of the spiky coating and on the base reactivity, is the
main interest of our study.  Qualitatively, $- z_0$ can be understood
as the location of an effective flat surface that produces the same
far-away concentration profile as the coated surface
\cite{Sarkar_1995,Vandembroucq_1997,Blyth_1993,Blyth_2003,Skvortsov_2014}:
\begin{equation}  \label{eq:kappa_eff}
\c \approx \c_0 \frac{z + z_0}{L_2 + z_0} \quad (z \geq -z_0),
\end{equation}
where $\c = \c_0$ at $z = L_2$ and $\c = 0$ at $z = - z_0$.
Alternatively, the effect of the coated surface can be represented by
a flat partially reactive surface at $z = 0$, with the effective
reactivity $\kappa_{\rm eff} = D/z_0$ and the effective boundary
condition $D\, \partial \c/\partial z = \kappa_{\rm eff}\, \c$ at $z =
0$.  The simple but universal form (\ref{eq:Jexact}) of the total flux
and the exact expression for $z_0$ are among the main results of the
paper.  Moreover, we will show that $z_0$ is (almost) independent of
$L_2$ whenever $L_2 \gg R_2$, but depends on two rescaled dimensions
of the pillar: $\rho = R_1/R_2$ and $h = L_1/R_2$.  We study the
behavior of the total flux in different asymptotic regimes; in
particular, we discover some new asymptotic relations and retrieve a
few known ones such as, e.g., trapping by a disk in a tube ($h = 0$).
We also analyze the effect of the reactivity of the base by
considering reflecting, absorbing and partially reactive cases.

The paper is organized as follows.  In Sec. \ref{sec:main}, we
formulate the mathematical problem, sketch the main steps of its exact
solution, present an approximation solution and discuss some general
properties.  Section \ref{sec:discussion} explores the dependence of
the total flux on the geometric parameters of the pillar.  Section
\ref{sec:conclusion} presents concluding remarks.  The mathematical
details of the derivation are re-delegated to Appendix.

\section{Main results}
\label{sec:main}

We consider point-like particles diffusing inside a three-dimensional
layer between parallel planes at $z = -L_1$ and $z = L_2$.  At steady
state, the particle concentration $\c$ obeys the three-dimensional
Laplace equation
\begin{equation}
\label{A0}
\Delta \c  = 0.
\end{equation}
The particle concentration at the top plane ($z=L_2$) is imposed to be
constant: $\c = \c_0$.  The bottom interface consists of an infinite
square lattice of absorbing cylindrical pillars of height $L_1$ and
radius $R_1$ protruding from a flat base at $z=-L_1$
(Fig. \ref{fig:scheme}(a)).  The surface of each pillar is absorbing
($\c = 0$), while the base can be absorbing, reflecting or partially
reactive.  To model spikes, one can set $R_1 \ll L_1$ but we treat the
general case.

Due to periodicity of the system in the $xy$ plane, one can consider
the solution of Eq. (\ref{A0}) within one periodic cell around one
pillar, i.e., within a cylinder $-L_1 \le z \le L_2$ with the square
cross-section $(-\ell/2,\ell/2) \times (-\ell/2,\ell/2)$ of the
lattice cell, with $\ell$ being the distance between centers of two
neighboring pillars (Fig. \ref{fig:scheme}(a)).  The symmetry of
the periodic cell allows one to replace periodic boundary conditions
on the square cross-section by reflecting (Neumann) boundary
condition.  In fact, eventual displacements between adjacent periodic
cells do not affect the diffusive flux from the top plane to the
absorbing coating.  Moreover, we apply the rationale from
\cite{Keller_1967} to replace the original periodic cell by a
circular tube of radius $R_2$ with the reflecting boundary condition
on its side wall (Fig. \ref{fig:scheme}(b)).  In this way, the
solution of the original problem is {\it approximated} by that of the
modified problem in a circular tube (see \cite{Bernoff18,Paquin22} and
references therein for other methods to treat planar problems on
Bravais lattices by preserving periodic conditions).  The radius $R_2$
is chosen to get the same cross-sectional area, i.e., to preserve the
volume of the cell; in the case of the square lattice, one gets
\begin{equation}  \label{eq:R2_ell}
R_2 = \frac{\ell}{\sqrt{\pi}}
\end{equation}
(similar constructions can be made for other regular lattices, e.g., a
hexagonal lattice).

From now on, we focus on steady-state diffusion inside a cylindrical
domain $\Omega$ of radius $R_2$, capped by parallel planes at $z =
-L_1$ and $z = L_2$, which contains another co-axial cylinder of
radius $R_1$ (an absorbing pillar), capped by parallel planes at $z =
-L_1$ and $z = 0$, as shown on Fig. \ref{fig:scheme}(b).  We search
for the harmonic function $u(r,z)$ satisfying the following boundary
value problem in cylindrical coordinates $(r,z,\phi)$:
\begin{subequations}  \label{eq:problem_u}
\begin{align}
& \Delta u = 0  \quad \textrm{in}~\Omega, \\  \label{eq:cond_top}
& u(r,L_2) = 1 \quad (r < R_2), \\  \label{eq:cond_disk}
& u(r,0) = 0 \quad (r < R_1),\\  \label{eq:cond_inner}
& u(R_1,z) = 0 \quad (-L_1 < z < 0),\\  \label{eq:cond_outer}
& (\partial_r u(r,z))_{r = R_2} = 0 \quad (-L_1 < z < L_2),  \\  \label{eq:cond_bottom}
& \bigl[D\partial_z u(r,z) + \kappa_b u(r,z)\bigr]_{z = -L_1} = 0 \quad (R_1 < r < R_2),
\end{align}
\end{subequations}
where $\Delta = \partial_r^2 + (1/r) \partial_r + \partial_z^2$ is the
Laplace operator in cylindrical coordinates (without the angular
part), $D$ is the diffusion coefficient, and $\kappa_b$ is the
reactivity of the base, which can take any value between $\kappa_b =
0$ for a reflecting base and $\kappa_b = \infty$ for an absorbing
base.  Here, Eq. (\ref{eq:cond_top}) sets the homogeneous source of
particles at the top boundary, Eqs. (\ref{eq:cond_disk},
\ref{eq:cond_inner}) incorporate the absorbing pillar,
Eq. (\ref{eq:cond_outer}) describes the reflecting outer cylindrical
surface, and Eq. (\ref{eq:cond_bottom}) accounts for partial
reactivity of the bottom base.  The rotation invariance of this
problem implies that $u$ does not depend on the angle $\phi$, which
therefore will be omitted in what follows.
Multiplication of $u(r,z)$ by $\c_0$ yields the concentration of
particles inside the domain, while
\begin{equation}
J = \c_0\int\limits_0^{R_2} dr \, r \, \bigl(D \partial_z u(r,z)\bigr)_{z=L_2}
\end{equation}
is the total diffusive flux of particles that enter into the domain
from the top disk.  Due to the conservation of particles inside the
domain, this is precisely the diffusive flux of particles onto the
absorbing boundary (i.e., the pillar and the reactive base if
$\kappa_b > 0$).
We aim at finding exact and approximate solutions of the problem
(\ref{eq:problem_u}) and then analyze the dependence of the total flux
$J$ on the geometric parameters $L_1$, $R_1$, $L_2$, $R_2$ for
different $\kappa_b$.

Note that $u(r,z)$ can be interpreted as the splitting probability: if
the particle started from a point $(r,z)$, then $u(r,z)$ is the
probability of its first arrival onto the top disk before hitting the
pillar (or reacting on the bottom base if $\kappa_b >0$).  This
interpretation opens a way to access $u(r,z)$ and the total flux by
Monte Carlo simulations, see Appendix \ref{sec:extensions} for
details.

\subsection{Exact solution}

The exact solution of Eq. (\ref{eq:problem_u}) can be found by a mode
matching method (see \cite{Delitsyn18,Grebenkov18,Delitsyn22} and
references therein).  In a nutshell, one can represent a harmonic
function $u(r,z)$ in the subdomains $-L_1 < z < 0$ and $0 < z < L_2$
as two series involving appropriate Bessel functions.  Matching these
two representations at $z = 0$ yields an infinite system of linear
algebraic equations on the unknown coefficients of these series.  The
last step entails solving this system of equations by truncation and
numerical inversion of a finite-size matrix with explicitly known
elements.  Despite the need for the numerical inversion of a matrix,
the resulting solution yields an analytical dependence of $u(r,z)$ on
$r$ and $z$, offers an efficient tool for its numerical computation,
and allows for asymptotic analysis.  Even though the exact solution is
one of the main results of this paper, we re-delegate the details of
its derivation to Appendix \ref{sec:exact}.  In particular, the total
diffusive flux $J$ is given by Eq. (\ref{eq:J}) that we rewrite with
the aid of Eq. (\ref{eq:c02_bis}) in the form (\ref{eq:Jexact}).  The
offset parameter $z_0$ depends on the reactivity $\kappa_b$ of the
base and on the geometric parameters $R_1$, $L_1$, $L_2$, $R_2$ in a
sophisticated way via the exact relation (\ref{eq:z0}).  This
dependence is the main object of our study.

\subsection{Diagonal approximation} 

The exact relation (\ref{eq:z0}) determining the offset parameter
$z_0$ requires the inversion of the infinite-dimensional matrix
$I+W'$, whose elements are known explicitly (see Appendix
\ref{sec:exact}).  As a consequence, the dependence of $z_0$ on the
geometric parameters remains hidden through this inversion, which has
to be done numerically.
Motivated by the ideas of Refs.
\cite{Benichou11,Calandre14,Benichou15}, we conducted a thorough
inspection of numerical results for different configurations of
parameters.  This revealed that the diagonal elements of the matrix
$W'$ are generally dominant as compared to non-diagonal elements.
This let us to propose a diagonal approximation, in which the matrix
$W'$ is replaced by a diagonal matrix formed by the diagonal elements
of $W'$.  In other words, all non-diagonal elements of this matrix are
set to $0$ that yields
\begin{equation}  \label{eq:Japp}
J_{\rm app} = \frac{\pi R_2^2 D \c_0}{L_2 + z_{0,\rm app}} \,,
\end{equation}
with
\begin{equation}  \label{eq:z0_app}
z_{0,\rm app} = L_2 \biggl(W_{0,0} - \sum\limits_{k=1}^\infty \frac{W_{0,k} W_{k,0}}{1 + W_{k,k}} \biggr),
\end{equation}
where the elements of the matrix $W$ (or $W'$) are given explicitly by
Eq. (\ref{eq:Wdef}).  Even though the dependence of the offset
$z_{0,\rm app}$ on the geometric parameters may still be complicated,
it is fully explicit within this approximation, as there is no matrix
inversion anymore.  As discussed in the following, this approximation
gives very accurate results for a broad range of geometric parameters.

\subsection{General properties of the rescaled offset}
\label{sec:general}

We aim at understanding how the offset parameter $z_0$ (and thus the
total flux $J$) depends on four dimensionless parameters:
\begin{equation}
\rho = \frac{R_1}{R_2} \,, \quad h = \frac{L_1}{R_2} \,, \quad H = \frac{L_2}{R_2} \,, \quad \kappa = \frac{\kappa_b}{D} R_2 .
\end{equation}
By construction, the rescaled radius $\rho$ of the pillar ranges from
$0$ to $1$, whereas its rescaled height $h$ can take any positive
value: $h \geq 0$.

At the end of Appendix \ref{sec:distant}, we show that the offset
$z_0$ does not almost depend on $L_2$ whenever $L_2 \gg R_2$, i.e.,
when the source of particles is much further than the inter-pillar
distance, which is a typical setting for most applications.  This is
an expected behavior because $z_0$ (or, equivalently, $\kappa_{\rm
eff} = D/z_0$) is the intrinsic ``material'' property of the spiky
coating (like an impedance), so it is a function of $R_1$, $R_2$,
$L_1$, and $\kappa_b/D$, but it should be independent of $L_2$.  In
the following, we impose the condition $L_2\gg R_2$ to eliminate the
geometric parameter $L_2$ from the analysis and rewrite
Eq. (\ref{eq:Jexact}) as
\begin{equation}  \label{eq:J_j}
J = \frac{\pi R_2^2 D\c_0}{L_2 + R_2 \zeta_\kappa(\rho,h)}  \,,
\end{equation}
where 
\begin{equation}  \label{eq:zeta}
\zeta_\kappa(\rho,h) = \lim\limits_{L_2\to\infty} \frac{z_0}{R_2}
\end{equation}
is the dimensionless offset in the limit $L_2\to\infty$.  One sees
that the total flux $J$ depends on $L_2$ in a trivial way, i.e., via
$L_2$ standing in the denominator of Eq. (\ref{eq:J_j}).  In turn, the
rescaled offset aggregates the dependences on other parameters.

Let us first look at the function $\zeta_{\kappa}(\rho,h)$ in the
limit $\rho = 1$ when the pillar fills the whole tube for $-L_1 < z <
0$, and one simply gets one-dimensional diffusion from the top disk to
the bottom absorbing disk, with the elementary solution $u_0(r,z) =
z/L_2$ and the total flux
\begin{equation}
J_{\rm max} = \frac{\pi R_2^2 D\c_0}{L_2} \,, 
\end{equation}
i.e.,
\begin{equation}  \label{eq:zeta_rho1}
\zeta_\kappa(1,h) = 0.
\end{equation}
In this trivial limit, there is no base (other than the pillar itself)
so that the flux does not depend neither on $\kappa$, nor on $h$.

In turn, if $\rho < 1$, thinning of the pillar removes an easily
accessible part of the absorbing disk at $z = 0$ with $R_1 < r <
R_2$ and thus diminishes the diffusive flux.  As a consequence, the
configuration with $\rho = 1$ yielded the maximal flux, i.e.,
\begin{equation}  \label{eq:ineq1}
1 \geq \frac{J}{J_{\rm max}} = \frac{1}{1 + \zeta_\kappa(\rho,h) \frac{R_2}{L_2}} 
\end{equation}
so that
\begin{equation}  
\zeta_\kappa(\rho,h) \geq 0
\end{equation}
for any set of parameters $\rho$, $h$ and $\kappa$.  In other words,
the offset is always positive.  This observation may sound
counter-intuitive because the total absorbing surface of the pillar,
$\pi R_1^2 + 2\pi R_1 L_1$, can dramatically increase due to its
cylindrical part (the second term).  However, this cylindrical part is
less accessible to Brownian particles due to the so-called diffusional
screening 
\cite{Traytak92,Felici03,McDonald03,McDonald04,Traytak07,Andrade07,Traytak18}
by the top disk (or by the singularity in concentration profile when
the diameter of the top disk tends to zero) so that the gain in the
total absorbing surface is compensated by its lower accessibility
\cite{Grebenkov05,Grebenkov05b}.

To get more insights into this property, one can rely on an equivalent
electrostatic problem of finding the electric current $J$ through a
conducting medium between two metallic electrodes (at the top and the
bottom), to which a voltage $\c_0$ is applied \cite{Liu85,Sapoval94}.
In the case of two flat parallel electrodes at $z = L_2$ and $z = 0$,
one has $z_0 = 0$ and Eq. (\ref{eq:Jexact}) yields the overall
resistance, $\mathcal{R}_0 = \c_0/J = L_2/(D \pi R_2^2)$, where $1/D$
can be interpreted as the resistivity of the medium, while $L_2$ and
$\pi R_2^2$ are the length and the cross-sectional area of such a
cylindrical wire.  In turn, if the conducting medium lies between the
top flat electrode and the bottom spiky electrode, the inclusion of
the bottom part with $z < 0$ adds more resistive material between two
electrodes and can only increase the overall resistance and thus
decrease the current, in agreement with the inequality
(\ref{eq:ineq1}).  In this electrostatic analogy, it is clear that an
increased area of the bottom electrode does not matter.  In this
light, Eq. (\ref{eq:Jexact}) can also be written as
\begin{equation}
J = \frac{\c_0}{\mathcal{R}_0 + \mathcal{R}_b} \,, \qquad \mathcal{R}_b = \frac{z_0}{\pi R_2^2 D} \,,
\end{equation}
where $\mathcal{R}_b$ can be interpreted as the resistance of the
lower layer $-L_1< z < 0$, which is connected in series to the
resistance $\mathcal{R}_0$ of the upper (flat) layer $0 < z < L_2$.
This representation provides yet another revealing insight onto $z_0$
as the length of an effective ``wire'' of the cross-sectional area
$\pi R_2^2$ and resistivity $1/D$.  Even though this electrostatic
analogy was formulated for the absorbing base, the underlying
arguments can be extended to the reflecting base as well.

In order to quantify the diffusional screening, in Appendix
\ref{sec:Atop} we computed the flux of particles onto the top of the
pillar, $J_{\rm top}$, which determines the fraction of particles,
$J_{\rm top}/J$, captured there.  Moreover, when the base is not
reflecting and can thus capture some particles, it is instructive to
get the fraction of particles absorbed by the base.  For this purpose,
we also computed the flux onto the base $J_{\rm base}$.  We will
discuss the behavior of these two quantities in the next section.
 
It is worth noting that some references reported negative values of
$z_0$ (see, e.g., \cite{Skvortsov_2019}).  However, this is a
consequence of convention that was used in that references to write
the total flux as $J = \pi R_2^2 D \c_0/(L + z'_0)$, where $L = L_1 +
L_2$ is the total distance between the top disk at $z = L_2$ and the
bottom base at $z = - L_1$, and prime distinguishes the new offset
$z'_0$ from our parameter $z_0$.  Evidently, one has $z'_0 = z_0 -
L_1$, i.e., $z'_0$ can take positive or negative values depending on
whether $z_0$ is larger or smaller than $L_1$.  We keep using our
convention, which follows from the exact solution of the problem.

Finally, we stress that the effective flat absorbing surface at $z =
-z_0$ may lie below the bottom base at $z = -L_1$, i.e., outside the
considered domain.  If one needs to avoid such locations, the lowest
location can be limited to the base level $z = -L_1$, but one would
have to replace the absorbing surface by partially reactive surface,
with the effective reactivity $\kappa'_{\rm eff}$ such that $z_0 = L_1
+ D/\kappa'_{\rm eff}$.  In other words, there are infinitely many
effective partially reactive surfaces at different locations, which
yield the same $z_0$ and thus the same total flux (see also
\cite{Martin_2022b}).

\section{Discussion}
\label{sec:discussion}

In this section, we investigate the dependence of the total flux
through the system on the geometric parameters in several asymptotic
regimes.  We start with a pillar on the reflecting base and then
consider the absorbing base.

\subsection{Reflecting base}

\begin{figure}
\begin{center}
\includegraphics[width=85mm]{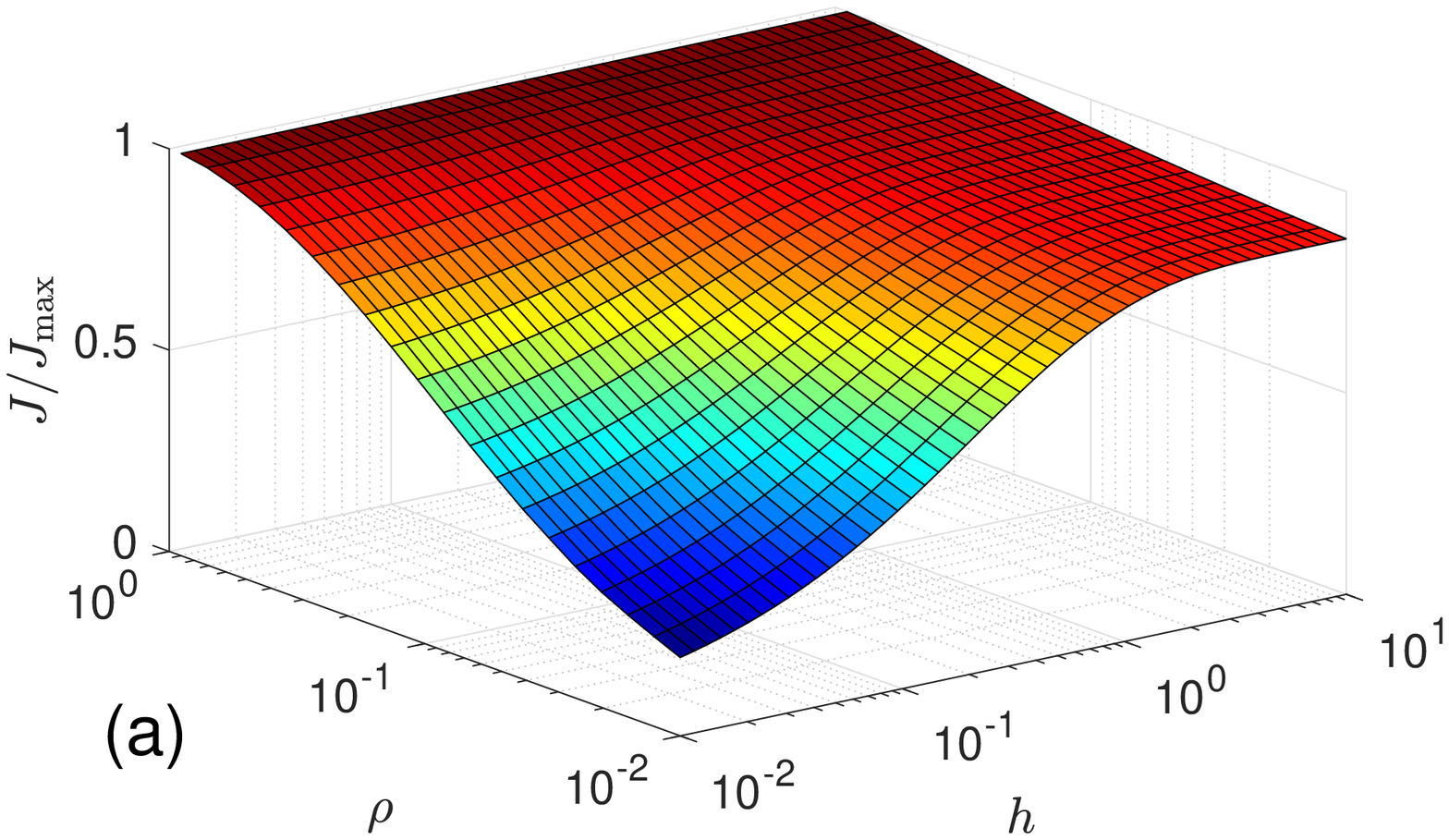} 
\includegraphics[width=85mm]{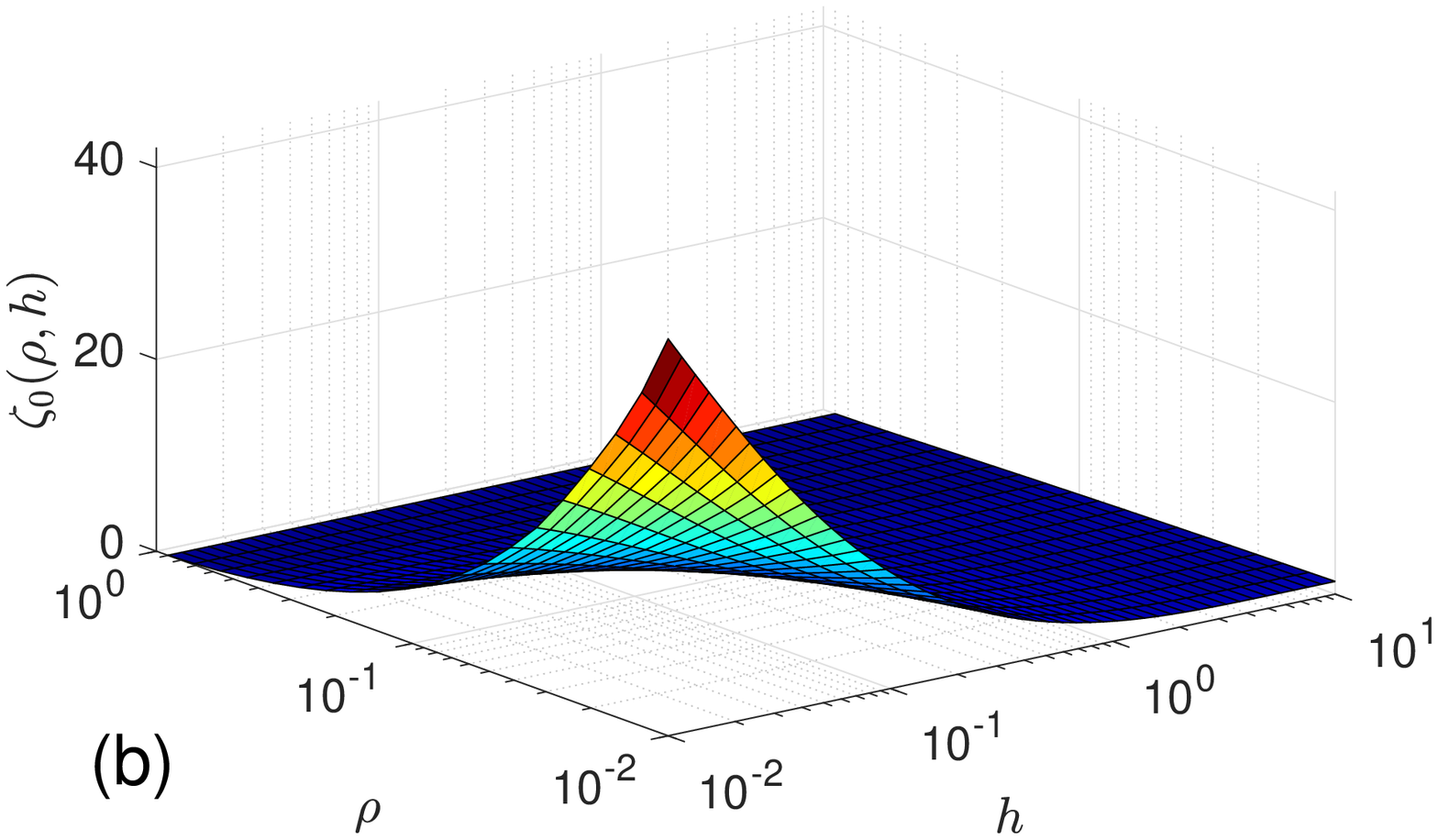} 
\includegraphics[width=85mm]{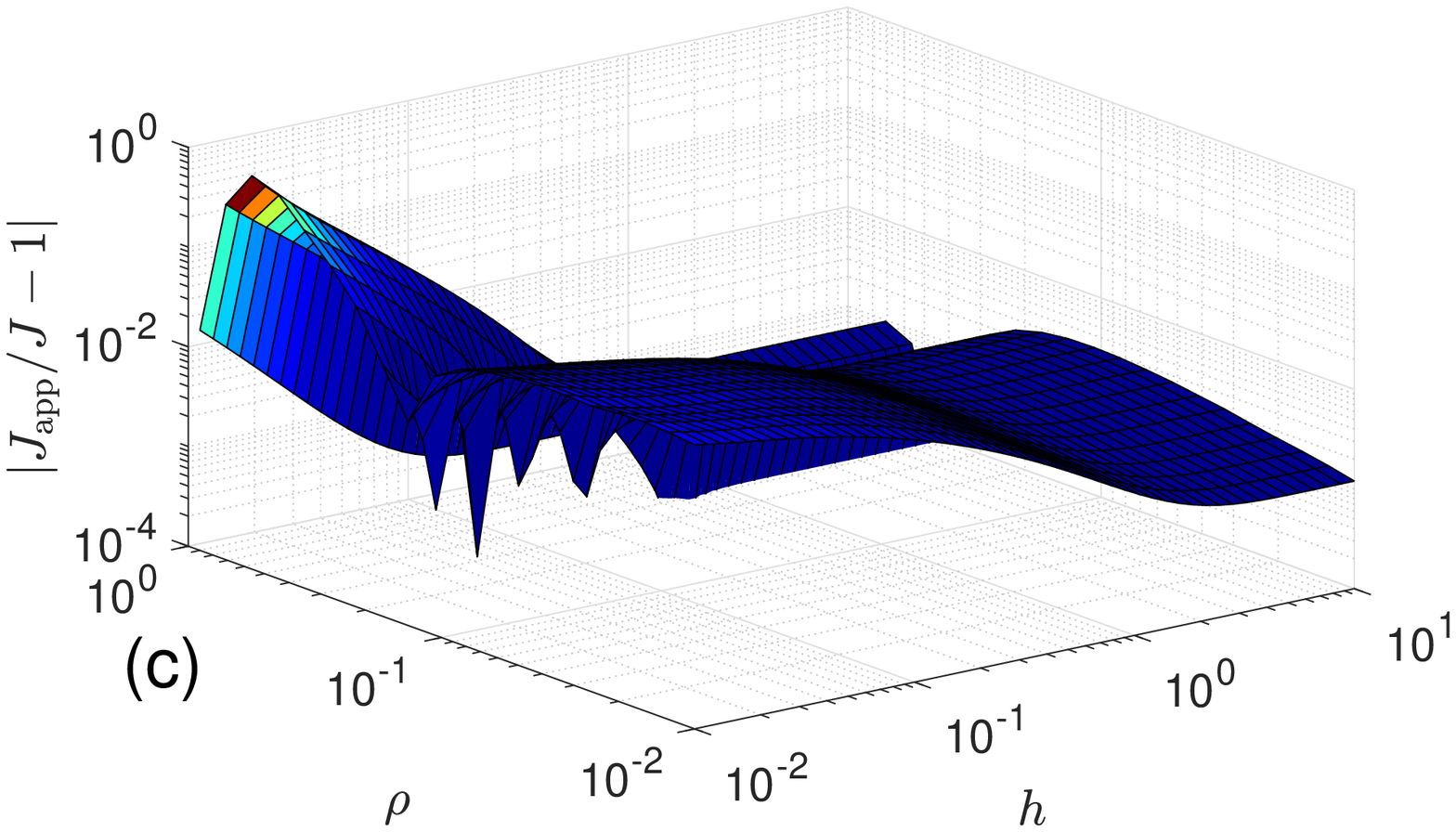} 
\end{center}
\caption{
{\bf (a)} The total flux $J$ (rescaled by $J_{\rm max} = \pi R_2^2
D\c_0/L_2$) as a function of $\rho = R_1/R_2$ and $h = L_1/R_2$ for
the reflecting base ($\kappa = 0$).  The function $J$ was computed
from Eq. (\ref{eq:J}), in which the coefficient $c_{0,2}$ was obtained
by inverting numerically the matrix $I+W$ truncated to the size
$N\times N$ with $N = 50$, see Appendix \ref{sec:exact} for details.
We used $L_2/R_2 = 10$.  {\bf (b)} The corresponding dimensionless
offset $\zeta_0(\rho,h)$ as a function of $\rho = R_1/R_2$ and $h =
L_1/R_2$.  {\bf (c)} The absolute value of the relative error,
$|J_{\rm app}/J - 1|$, of the diagonal approximation $J_{\rm app}$
from Eq. (\ref{eq:Japp}).}
\label{fig:Jsurf_kappa0}
\end{figure}

When the base is reflecting ($\kappa = 0$), the particles can only be
absorbed on the pillar.  Expectedly, the flux diminishes as the
absorbing pillar gets smaller, i.e., when either $\rho$ or $h$
decreases (or both).  Note that the flux vanishes in the limit $\rho
\to 0$ for any fixed $h$ because the pillar shrinks to a
one-dimensional interval (a needle), which is ``invisible'' for
Brownian motion \cite{Morters}.  In turn, the flux remains strictly
positive in the limit $h\to 0$ for any fixed $\rho > 0$ when the
pillar transforms into a planar absorbing disk, which remains
accessible for Brownian motion.  Panel (a) of
Fig. \ref{fig:Jsurf_kappa0} illustrates these properties of the total
flux, while the panel (b) presents the corresponding offset
$\zeta_0(\rho,h)$.  As stated earlier, the ratio $J/J_{\rm max}$ is
equal to $1$ for $\rho = 1$ and any $h$, and it remains below $1$ for
other parameters.  Moreover, one observes that $J$ and
$\zeta_0(\rho,h)$ are monotonous functions of $\rho$ and $h$.  Even
though this observation is intuitively expected, its rigorous
demonstration presents an interesting perspective for future work.

Let us now inspect the panel (c) of Fig. \ref{fig:Jsurf_kappa0}, which
shows the absolute value of the relative error, $|J_{\rm app}/J - 1|$,
of the diagonal approximation of the flux $J_{\rm app}$ given by Eq.
(\ref{eq:Japp}).  When the pillar is long enough (say, $h\gtrsim 1$),
the relative error is below $1\%$, whatever the rescaled radius $\rho$
is.  As the pillar gets shorter, the relative error increases, even
though the approximation remains applicable when $\rho$ is small
enough (e.g., the relative error is around $10\%$ when $\rho \lesssim
0.1$).  The worst case corresponds to short wide pillars (i.e., when
the pillar is close to a disk).  In this particular case, one can
employ another approximation, as discussed below.  Overall, we
conclude that the diagonal approximation (\ref{eq:Japp}) yields very
accurate results for a broad range of geometric parameters.  At the
same time, the dependence of the approximate offset $z_{0, \rm app}$
in Eq. (\ref{eq:z0_app}) on the geometric parameters yet requires
further analysis.  To shed light on it, we explore several asymptotic
regimes.

\subsubsection*{Disk-like pillar}

\begin{figure}
\begin{center}
\includegraphics[width=85mm]{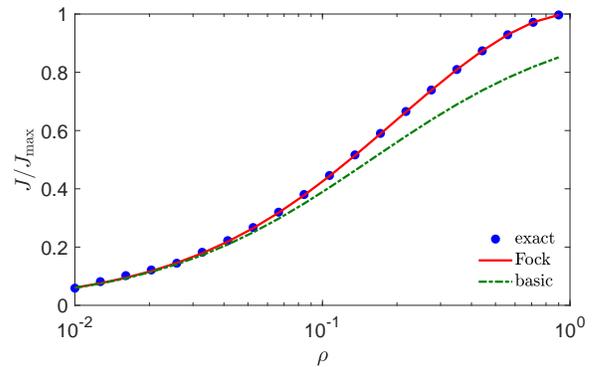} 
\end{center}
\caption{
The total flux $J$ (rescaled by $J_{\rm max} = \pi R_2^2 \c_0 D/L_2$)
as a function of $\rho = R_1/R_2$, at $L_2/R_2 = 5$, $L_1 = 0$ and
$\kappa = 0$ (an absorbing disk on the reflecting base).  Filled
circles present the exact solution obtained with the truncation order
$N = 50$, while solid line shows the flux from Eq. (\ref{eq:Jexact}),
with $z_0/R_2 = \zeta_0(\rho,0)$ given by Eq. (\ref{eq:Fock}).  Dashed
line presents the flux with $\zeta_0(\rho,0) \approx \pi/(4\rho)$,
i.e., without the Fock function $M(\rho)$.}
\label{fig:J_h0}
\end{figure}

We start by considering the limit $L_1 = 0$ when the pillar shrinks to
an absorbing disk of radius $R_1$ located on the reflecting base at $z
= 0$.  This problem was first studied by Fock \cite{Fock41} and later
by other researchers (see \cite{Skvorstov21,Grebenkov22c} and
references therein).  In particular, an approximate analytical
expression for the steady-state diffusive flux has been derived for a
tube that contains a perpendicular barrier with a circular hole
\cite{Skvorstov21}.  This setting is identical to the case of a
disk-like pillar on the reflecting base, so that we should get
\begin{equation}  \label{eq:Fock}
\zeta_0(\rho,0) \approx \frac{\pi}{4\rho} M(\rho),
\end{equation}
where $M(\rho)$ is the Fock function, which can be
approximated as \cite{Skvorstov21,Grebenkov22c} (see also
\cite{Leppington_1972}):
\begin{equation}  \label{eq:M_Fock}
M(\rho) = \frac{(1-\rho^2)^2}{1 + 1.37 \rho - 0.37 \rho^4} \,.
\end{equation}
Figure \ref{fig:J_h0} confirms the remarkable accuracy of the total
flux given by Eq. (\ref{eq:Jexact}) with $z_0 = R_2 \zeta_0(\rho,0)$
from Eq. (\ref{eq:Fock}) over the whole range of radii $\rho$.  Note
that a simpler approximation $\zeta_0(\rho,0) \approx
\pi/(4\rho)$ (i.e., without the Fock function) is still accurate at
small $\rho$ but exhibits moderate deviations at larger $\rho$.

\subsubsection*{Capacitance approximation}

Before dwelling into the asymptotic analysis of the exact solution, it
is instructive to discuss the capacitance approximation, which is
commonly used for estimating the steady-state flux and other related
quantities
\cite{Redner_2001,Krapivsky,Hughes,Zhou94,Grebenkov20}.
When both dimensions of the pillar ($R_1$ and $L_1$) are much smaller
than $R_2$ and $L_2$, it is tempting to treat the outer reflecting
cylindrical surface and the upper absorbing disk as being at infinity,
i.e., to consider a single absorbing pillar in the infinite upper
half-space (above the reflecting base).  In this case, the total flux
is determined by the capacitance $\C$ of the absorbing pillar as
$J_{\infty} = \c_0 D \C$, where $\c_0$ is the concentration kept at
infinity (here we employ the definition, according to which the
capacitance of a sphere of radius $R$ is $4\pi R$).  The mirror
reflection of the pillar with respect to the reflecting base allows
one to express $\C$ as half of the capacitance $\C_0$ of the twice
longer pillar in the whole space $\R^3$.  The latter is known exactly
\cite{Sandua_2013} and thus yields
\begin{equation}  \label{eq:Cpillar}
\C = \frac12 \C_0 = 2\pi R_1 \frac{1 + (L_1/R_1)^2}{\frac{\pi}{2} + \frac{L_1}{R_1} \ln \frac{L_1}{R_1}} 
\end{equation}
(note that former expressions for the capacitance $\C_0$ were given in
terms of a series expansion in powers of $1/\ln(L_1/R_1)$, see e.g.,
\cite{Jackson00,Chang70,Riba18} and references therein).  As $L_1\to
0$, one retrieves half of the capacitance $8R_1$ of an absorbing disk
in $\R^3$.  The factor $1/2$ in the above relation corrects the twice
larger total flux, which was artificially doubled by mirror reflection
with respect to the reflecting base.
The total flux onto the pillar can thus be approximated as
\begin{equation}  \label{eq:Jinfty}
J_{\infty} = \c_0 D \C  = 2\pi \c_0 D R_1 \frac{1 + (L_1/R_1)^2}{\frac{\pi}{2} + \frac{L_1}{R_1} \ln \frac{L_1}{R_1}} \,.
\end{equation}

Strictly speaking, this approximation is incompatible with our exact
form (\ref{eq:Jexact}), which is valid for any {\it finite} $R_2$ and
$L_2$.  In fact, when $R_2$ is fixed and $L_2$ increases,
Eq. (\ref{eq:Jexact}) implies that $J \propto 1/L_2$, i.e., the total
flux vanishes as $L_2\to \infty$, in contradiction with
Eq. (\ref{eq:Jinfty}).  In other words, the approximation
(\ref{eq:Jinfty}) might only be applicable in the double limit when
$R_2\to \infty$ and $L_2\to\infty$ simultaneously.  However, there are
infinitely many ways to relate $R_2$ and $L_2$ in order to undertake
this double limit.  While the detailed inspection of this mathematical
issue is beyond the scope of the paper, we mention one possible way.

One can rely on earlier works (see, e.g.,
\cite{Berezhkovskii_2004,Berezhkovskii07} and references therein) that
expressed the effective reactivity of an absorber inside a reflecting
tube as $\kappa_{\rm eff} \approx D \C/{\mathcal A}$, where ${\mathcal
A} = \pi R_2^2$ is the surface area of the tube cross-section.  As a
consequence, the total flux is given by Eq. (\ref{eq:Jexact}), with
\begin{equation}
z_0 = \frac{D}{\kappa_{\rm eff}} \approx \frac{\pi R_2^2}{\C}   \,.
\end{equation}
Substituting Eq. (\ref{eq:Cpillar}), we get
\begin{equation}  \label{eq:zeta_capacitance}
\zeta_0(\rho,h) \approx \biggl(2\rho \, \frac{1 + (h/\rho)^2}{\frac{\pi}{2} + \frac{h}{\rho} \ln \frac{h}{\rho}}\biggr)^{-1} 
\end{equation}
for $\rho \ll 1$ and $h \ll 1$.  In the double limit $L_2\to \infty$
and $R_2\to \infty$ with $H = L_2/R_2$ being fixed, the total flux
tends to the limit $J_\infty$ from Eq. (\ref{eq:Jinfty}) for the
pillar in the upper half-space, which is independent of $H$, as
expected.  However, the quality of this approximation for finite $L_2$
and $R_2$ depends on $H$.  We stress that the limit $R_2\to\infty$ is
singular; in particular, the spectrum of the Laplace operator in the
unbounded domain with $R_2 = \infty$ is not discrete anymore, and
other mathematical tools are required to analyze the asymptotic
behavior.  For this reason, the above discussion remains speculative
from the mathematical point of view.

\begin{figure}
\begin{center}
\includegraphics[width=85mm]{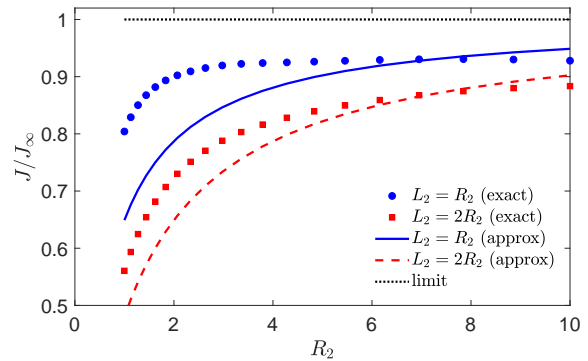} 
\end{center}
\caption{
The total flux $J$ (rescaled by $J_{\infty}$) as a function of $R_2$
for $R_1 = 0.1$, $L_1 = 0.5$, and two choices $L_2 = R_2$ (circles)
and $L_2 = 2R_2$ (squares).  Symbols present the exact solutions with
the truncation order $N = 50$.  Lines show Eq. (\ref{eq:Jexact}) with
$z_0 = R_2 \zeta_0(\rho,h)$ given by the approximation
(\ref{eq:zeta_capacitance}).  Dotted horizontal line indicates the
limit determined by the flux $J_\infty$ from Eq. (\ref{eq:Jinfty}),
which corresponds to the double limit $R_2 \to \infty$ and $L_2
\to\infty$.  }
\label{fig:Jinf}
\end{figure}

Figure \ref{fig:Jinf} illustrates the asymptotic behavior by showing
how the total flux increases when both $R_2$ and $L_2$ increase with
$H = L_2/R_2$ being fixed.  In this double limit, two opposite effects
seem to almost compensate each other: on the one hand, an increase of
$L_2$ diminishes the flux (as the source is getting further from the
absorbing pillar); on the other hand, an increase of $R_2$ enlarges
the area of the source (the top disk) and thus the flux.  Even though
the capacitance approximation (\ref{eq:zeta_capacitance}) shows a
similar trend, it is not accurate.

We expect that the capacitance approximation may be deduced from the
asymptotic analysis of $\zeta_0(\rho,h)$ as both $\rho$ and $h$ go to
$0$, with $h/\rho= L_1/R_1$ being fixed.  However, even the numerical
computation of the total flux at very large $R_2$ is difficult because
larger and larger truncation orders are needed as $R_2$ increases.
Further theoretical analysis of this limit presents therefore an
interesting perspective for future research.

At the same time, we recall that very large $R_2$ corresponds to an
arrangement of very distant short pillars, which is not a common
setting for applications.  In the rest of the text, we keep $R_2$
fixed and consider $L_2 \gg R_2$, as stated earlier.

\subsubsection*{Thin pillar approximation}

\begin{figure}
\begin{center}
\includegraphics[width=85mm]{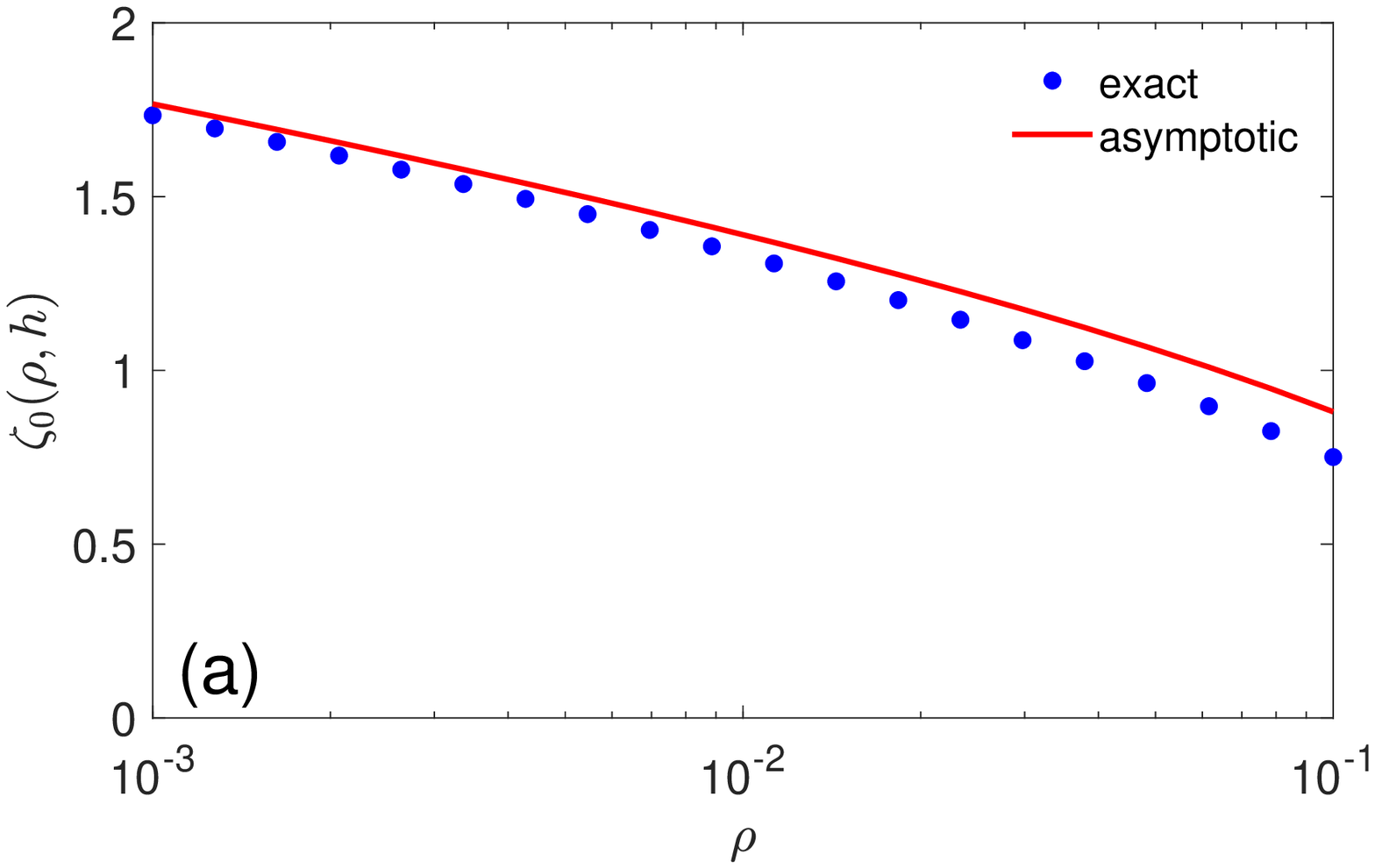} 
\includegraphics[width=85mm]{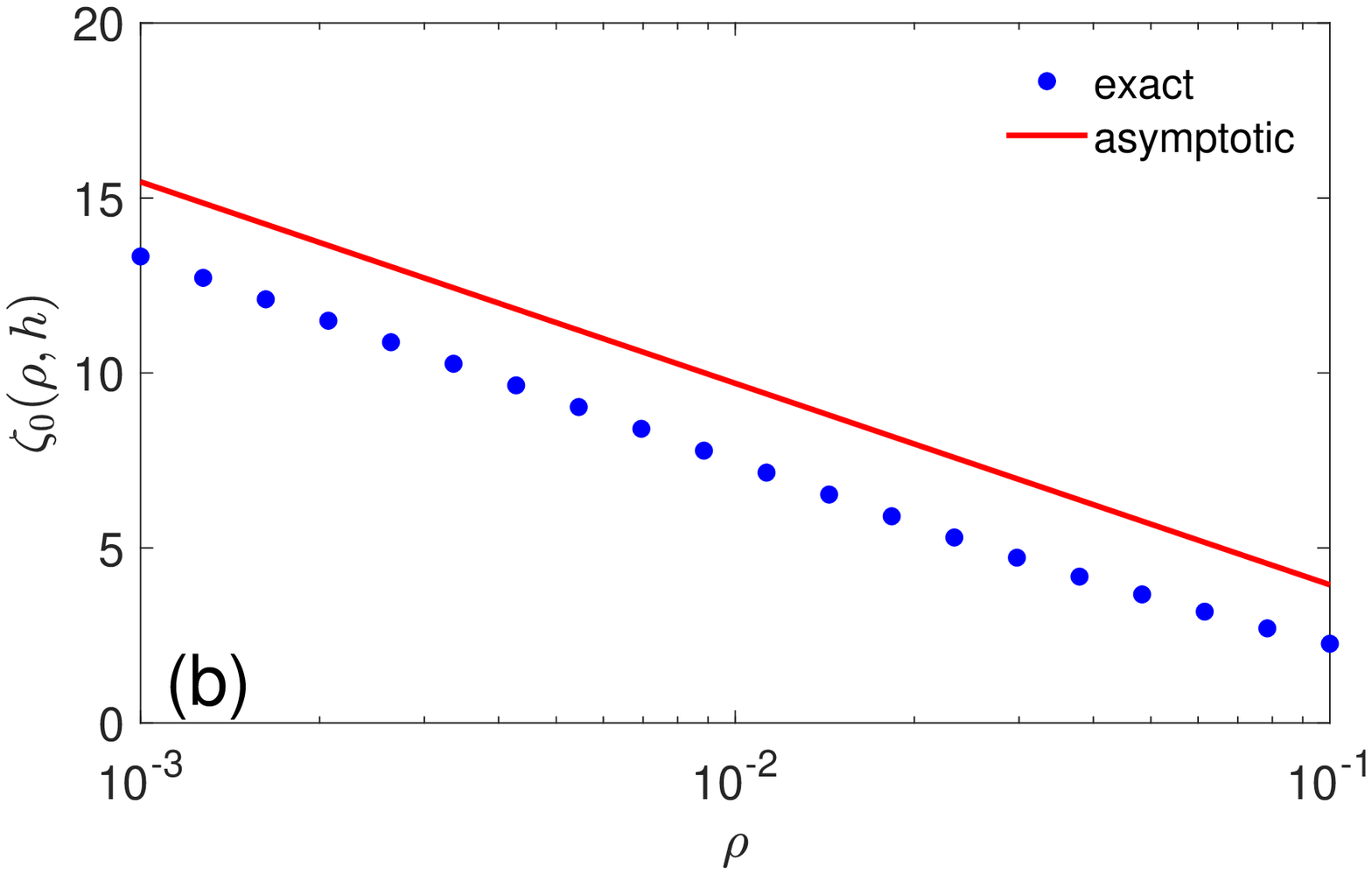} 
\end{center}
\caption{
The dimensionless offset $\zeta_0(\rho,h)$ as a function of $\rho$ for
$h = 5$ {\bf (a)} and $h = 0.2$ {\bf (b)} for the reflecting base
($\kappa = 0$).  Filled circles present this function deduced from for
the total flux, which was obtained by truncation at $N = 20$.  Solid
line shows the asymptotic relation (\ref{eq:zeta_rho0_kappa0}).}
\label{fig:zeta_rho}
\end{figure}

Let us now look at the setting $\rho \ll 1$, which corresponds to thin
pillars, which are well-separated from each other (i.e. $R_1 \ll \ell
= \sqrt{\pi} R_2$).  In this limit, the widths $R_2$ and $R_2 - R_1$
of the upper and lower subdomains are almost equal (see
Fig. \ref{fig:scheme}(c)), and they differ essentially by the boundary
condition on the left edge.  The absorbing boundary in the lower
subdomain can be treated via standard ``strong perturbation''
asymptotic tools \cite{Mazya85,Ward93}.  Re-delegating the derivation
steps to Appendix \ref{sec:Athin}, we discuss here the derived
asymptotic approximation:
\begin{equation}  \label{eq:zeta_rho0}
\zeta_\kappa(\rho,h) \approx \frac{1}{\alpha_{0,1}} \, \frac{\alpha_{0,1} + \kappa \tanh(\alpha_{0,1} h)}
{\kappa + \alpha_{0,1} \tanh(\alpha_{0,1} h)}  \quad (\rho \to 0),
\end{equation}
where 
\begin{equation}  \label{eq:alpha01}
\alpha_{0,1} \approx \frac{\sqrt{2}}{\sqrt{\ln(1/\rho) - 3/4}}    \qquad(\rho \to 0).
\end{equation}
For the reflecting base ($\kappa = 0$), one gets an even simpler
relation:
\begin{equation}  \label{eq:zeta_rho0_kappa0}
\zeta_0(\rho,h) \approx \frac{\ctanh(\alpha_{0,1} h)}{\alpha_{0,1}}    \quad (\rho \to 0),
\end{equation}
i.e., the offset logarithmically diverges as $\rho \to 0$, implying
that the total flux $J$ vanishes in this limit, as expected.  However,
as the offset $z_0 \approx R_2 \zeta_0(\rho,h)$ is added to a large
distance $L_2$, this decay is extremely slow, especially for large
$h$, as one can see on Fig. \ref{fig:Jsurf_kappa0}(a).  We stress that
the height of the pillar affects this speed: 

(i) if $h \gg 1$ (or even $h = \infty$), the argument $\alpha_{0,1} h$
can be large even for extremely small $\rho$, and the numerator of
Eq. (\ref{eq:zeta_rho0_kappa0}) can be replaced by $1$, yielding
\begin{equation}   \label{eq:zeta_rho0_kappa0_hinf}
\zeta_0(\rho,h) \approx \frac{\sqrt{\ln(1/\rho) - 3/4}}{\sqrt{2}}    \quad (\rho \to 0, ~ h\gg 1),
\end{equation}
independently of the height $h$.  Qualitatively, a very large height
of the pillar almost compensates for its vanishing radius.  

(ii) In contrast, if $0 < h \ll 1$, one gets
\begin{equation}   \label{eq:zeta_rho0_kappa0_small}
\zeta_0(\rho,h) \approx \frac{\ln(1/\rho) - 3/4}{2h}    \quad (\rho \to 0, ~ 0<h\ll 1).
\end{equation}
In this regime, the value of $\zeta_0(\rho,h)$ is also affected by the
height $h$: as the pillar is getting shorter, $\zeta_0(\rho,h)$
increases, as expected.  Note that the case $h = 0$ (an absorbing
disk) should be treated separately, as we did earlier.
Comparing the asymptotic expressions (\ref{eq:Fock},
\ref{eq:zeta_rho0_kappa0_small}), one sees that the order of the
limits, $\rho\to 0$ and $h\to 0$, is important.  This is a
manifestation of the absorber anisotropy, which was earlier reported
for other geometric settings (see, e.g.,
\cite{Grebenkov_2017,Chaigneau22}).

Figure \ref{fig:zeta_rho} illustrates the behavior of the offset
$\zeta_0(\rho,h)$ and its asymptotic approximation
(\ref{eq:zeta_rho0_kappa0}) for a long pillar ($h = 5$) and a short
pillar ($h = 0.2$).  In both cases, Eq. (\ref{eq:zeta_rho0_kappa0})
captures correctly the leading-order, divergent behavior.  In turn,
one can observe a nearly constant deviation (i.e., a shift between two
curves along the vertical axis), especially for the shorter pillar.
This deviation is caused by the next-order corrections which seem to
be of the order $O(1)$.  Further investigation of this asymptotic
behavior presents an interesting perspective of this work.

\subsubsection*{Diffusional screening by the top of the pillar}

\begin{figure}
\begin{center}
\includegraphics[width=85mm]{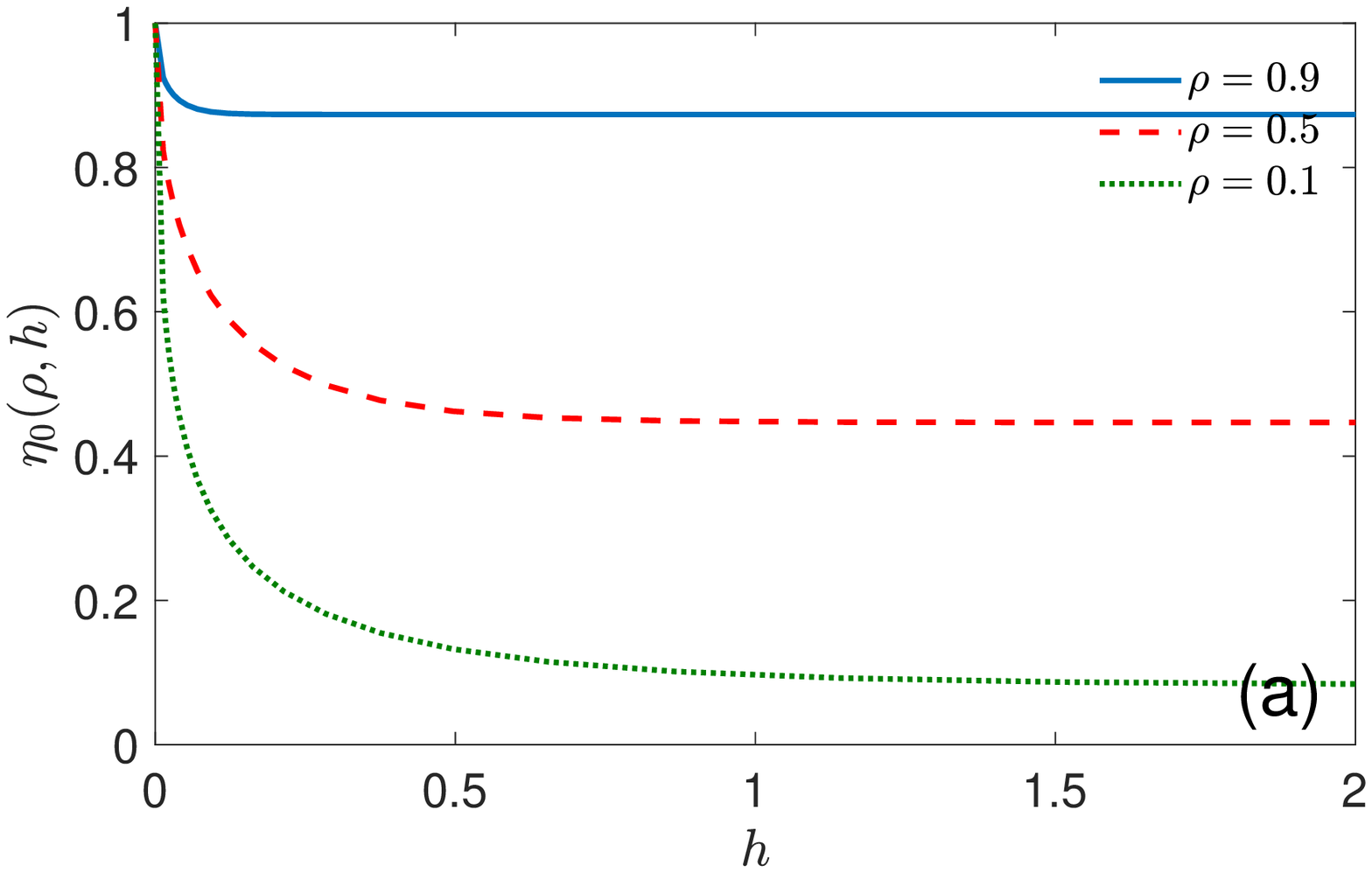} 
\includegraphics[width=85mm]{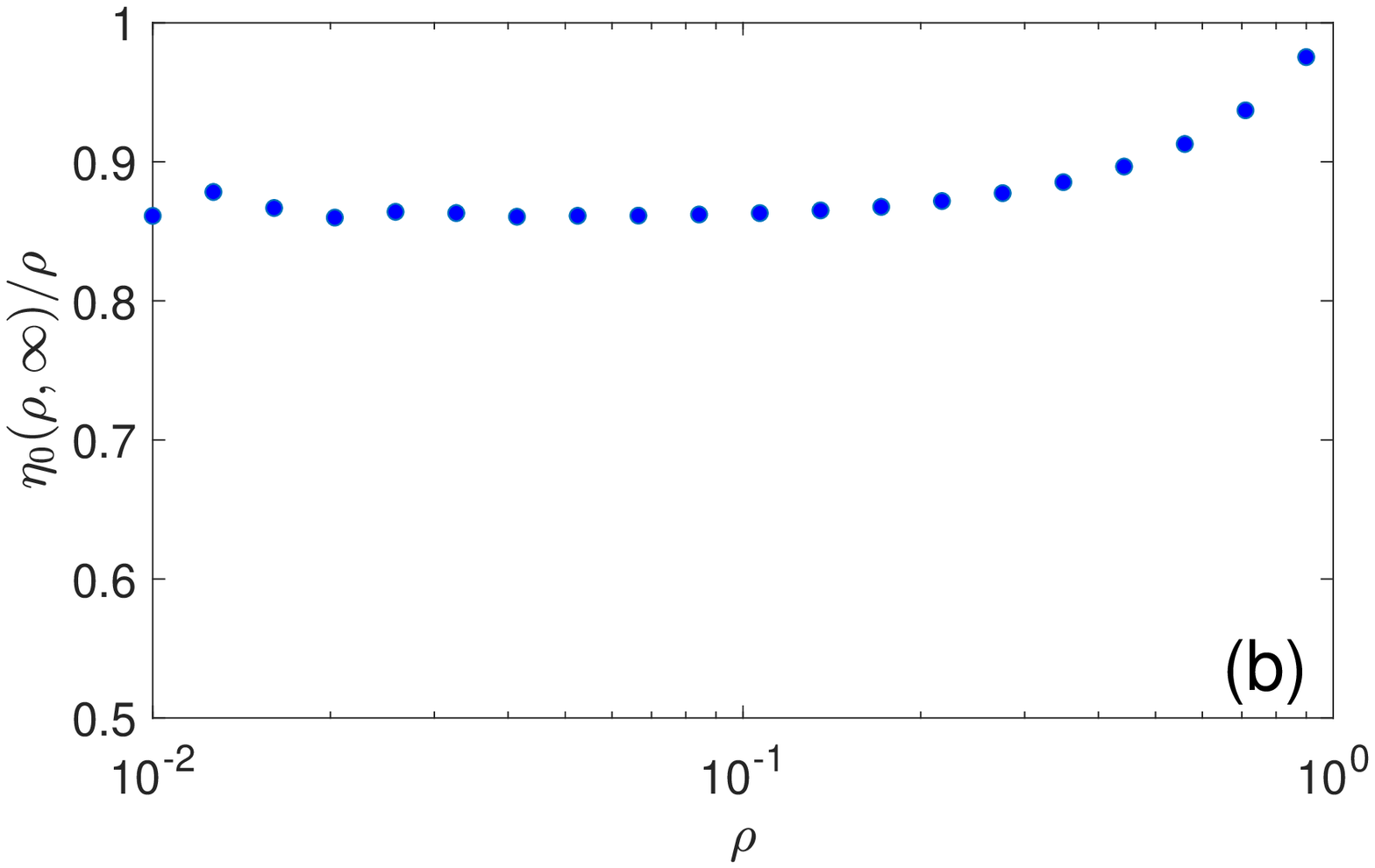} 
\end{center}
\caption{
{\bf (a)} The fraction $\eta_0(\rho,h)$ of particles absorbed on the
top of the pillar on the reflecting base ($\kappa = 0$), as a function
of the pillar height $h = L_1/R_1$, for three values of $\rho =
R_1/R_2$ (see the legend).  This fraction was obtained from
Eq. (\ref{eq:eta}) with $L_2/R_2 = 5$ and $N = 50$.  {\bf (b)} The
fraction $\eta_0(\rho,\infty)$ divided by $\rho$. }
\label{fig:Jtop}
\end{figure}

We complete this section by quantifying the diffusional screening
effect.  In Appendix \ref{sec:Atop}, we calculated the flux on the top
of the pillar, $J_{\rm top}$, and deduced the fraction of particles
absorbed there, $\eta_\kappa(\rho,h)$, as the limit of the ratio
$J_{\rm top}/J$ as $L_2\to \infty$, see Eq. (\ref{eq:eta}).  

Figure \ref{fig:Jtop}(a) illustrates the dependence of the ratio
$\eta_0(\rho,h)$ on the geometric parameters of the pillar: rescaled
height $h$ and radius $\rho$.  In the limit $h = 0$, there is no
cylindrical part of the pillar, and the whole flux is absorbed on the
top: $\eta_0(\rho,0) = 1$.  As $h$ increases, the cylindrical part of
the pillar starts to absorb the particles, and the ratio
$\eta_0(\rho,h)$ decreases to a strictly positive limit
$\eta_0(\rho,\infty)$ as $h\to \infty$.  This limiting value depends
on the radius of the pillar.  As $\rho$ stands in Eq. (\ref{eq:eta})
as a prefactor, we plot $\eta_0(\rho,\infty)/\rho$ on
Fig. \ref{fig:Jtop}(b).  One can see that this function grows from
$0.86$ to $1$.  It is worth noting, however, that an accurate
computation of $\eta_0(\rho,\infty)$ was difficult due to a slow
convergence of the series in Eq. (\ref{eq:Jtop}).  To overcome this
difficulty, we computed the ratio $J_{\rm top}/J$ from
Eq. (\ref{eq:Jtop}) by truncating matrices at orders $N = 100, 200,
\ldots, 1000$ and then performed a linear regression of the obtained
values versus $1/\sqrt{N}$ to extrapolate the ratio to the limit $N\to
\infty$.  Despite this refined analysis, one can notice minor
fluctuations of $\eta_0(\rho,\infty)/\rho$ at small $\rho$.  In
particular, we cannot rigorously state whether this ratio converges a
strictly positive constant around $0.86$ (as it seems to be the case
according to Fig. \ref{fig:Jtop}(b)), or exhibits an extremely slow
decay as $\rho \to 0$.  Further mathematical and numerical analysis of
this issue presents an interesting perspective.

\subsection{Absorbing base}
\label{sec:absorbing}

\begin{figure}
\begin{center}
\includegraphics[width=85mm]{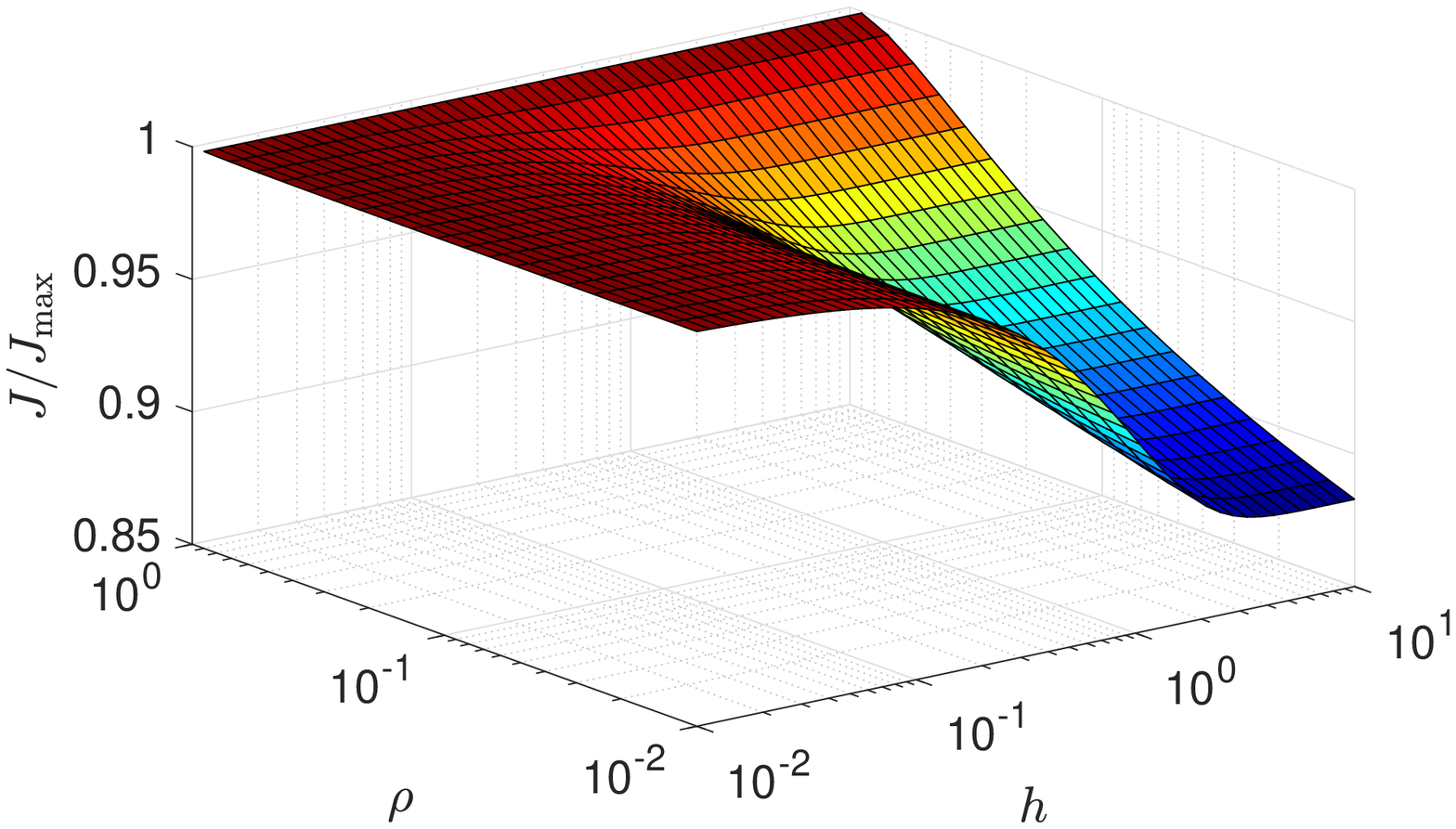} 
\includegraphics[width=85mm]{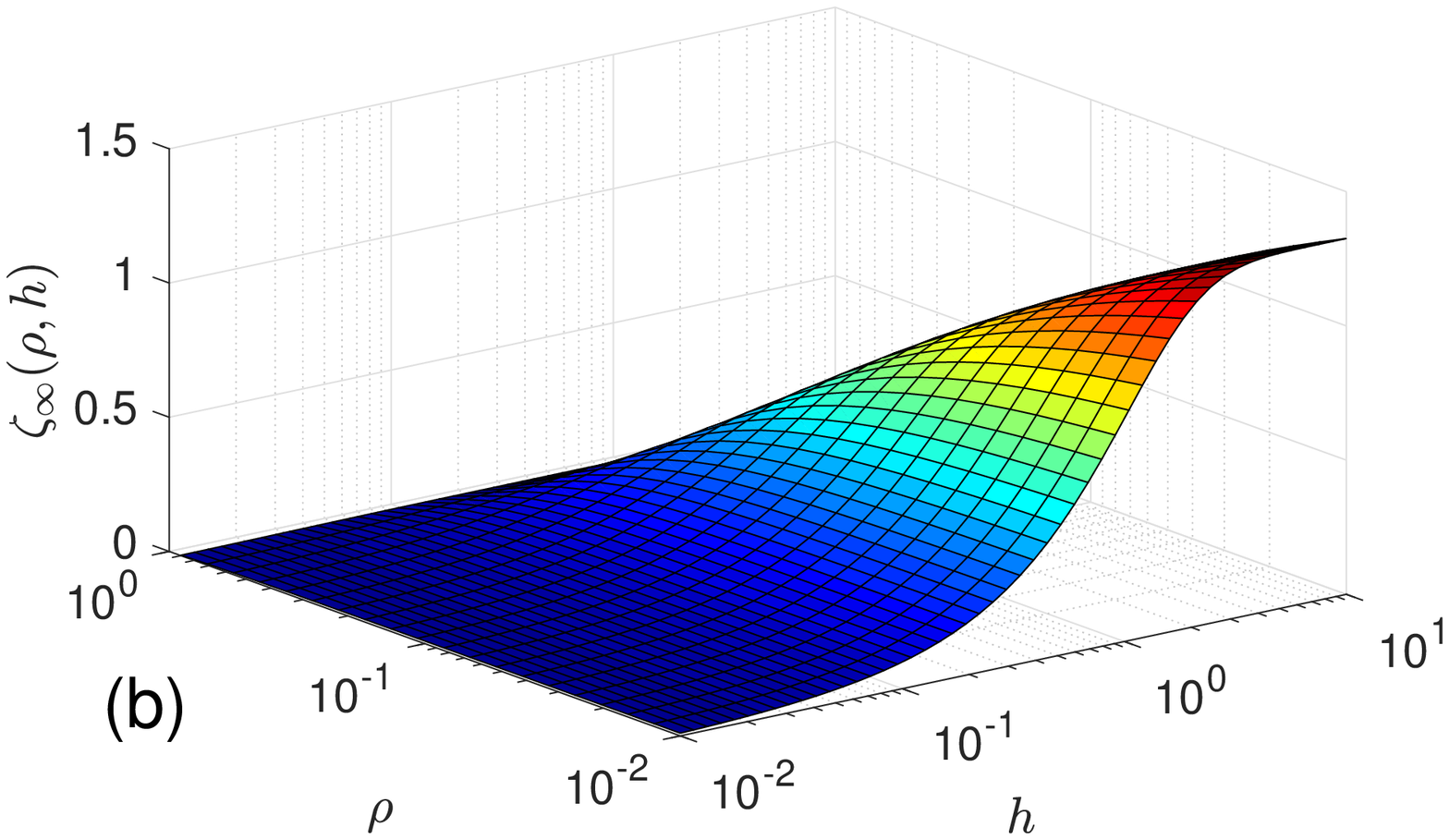} 
\includegraphics[width=85mm]{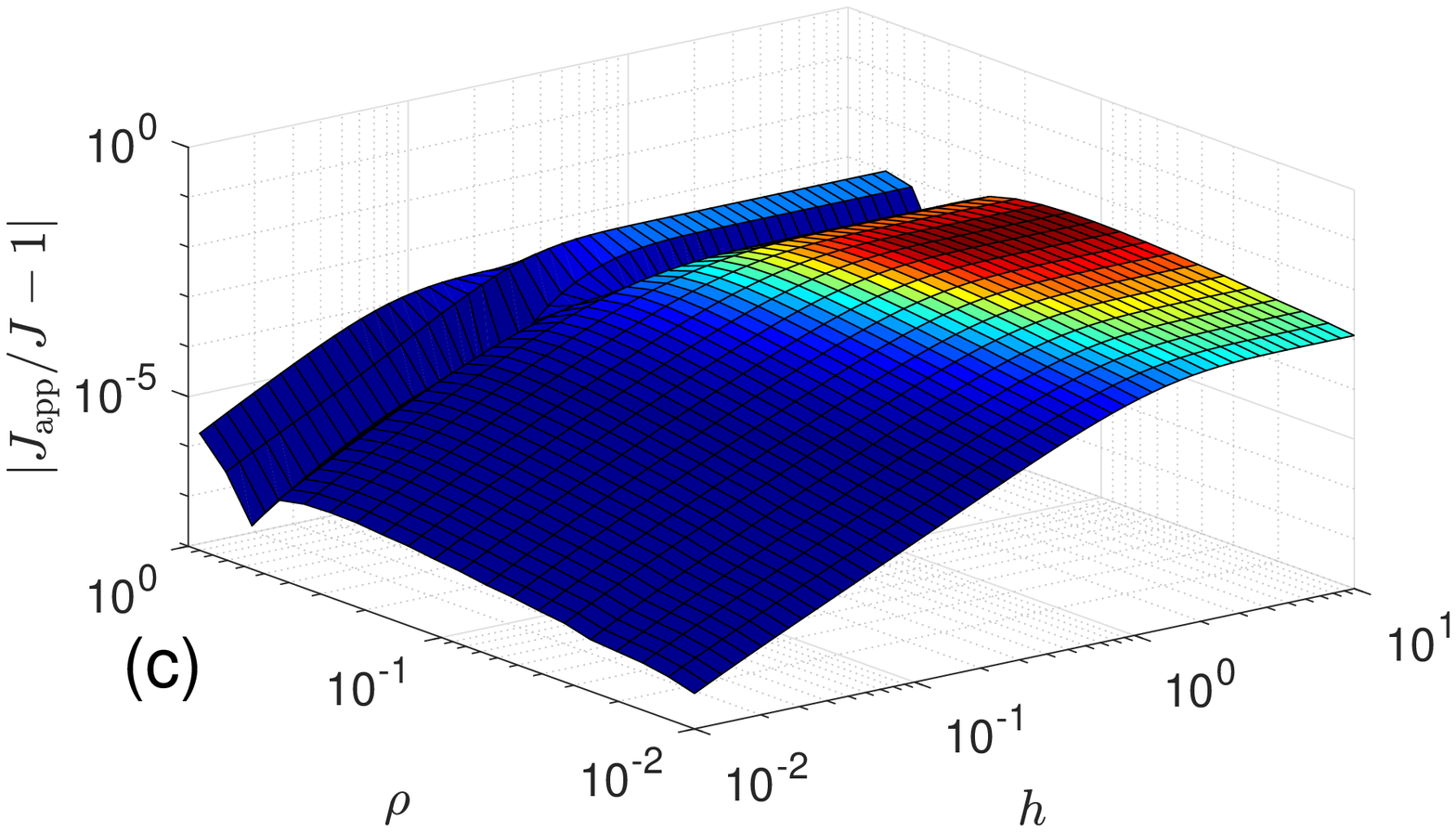} 
\end{center}
\caption{
{\bf (a)} The total flux $J$ (rescaled by $J_{\rm max} = \pi R_2^2
D\c_0/L_2$) as a function of $\rho = R_1/R_2$ and $h = L_1/R_2$ for
the absorbing base ($\kappa = \infty$).  The function $J$ was computed
from Eq. (\ref{eq:J}), in which the coefficient $c_{0,2}$ was obtained
by inverting numerically the matrix $I+W$ truncated to the size
$N\times N$ with $N = 50$, see Appendix \ref{sec:exact} for details.
We used $L_2/R_2 = 10$.  {\bf (b)} The corresponding dimensionless
offset $\zeta_\infty(\rho,h)$ as a function of $\rho = R_1/R_2$ and $h
= L_1/R_2$.  {\bf (c)} The absolute value of the relative error,
$|J_{\rm app}/J - 1|$, of the diagonal approximation $J_{\rm app}$
from Eq. (\ref{eq:Japp}).}
\label{fig:Jsurf_kappainf}
\end{figure}

Let us switch to the analysis of the absorbing pillar on the absorbing
base ($\kappa_b = \infty$).  As both the pillar and the base can now
absorb particles, the dependence of the total flux on the geometric
parameters is quite different.  Figure \ref{fig:Jsurf_kappainf}(a)
illustrates this dependence, while the panel (b) presents the
corresponding offset parameter $\zeta_\infty(\rho,h)$.  In addition to
the trivial limit (\ref{eq:zeta_rho1}) at $\rho = 1$, one also has
$\zeta_\infty(\rho,0) = 0$.  In fact, if the pillar shrinks to an
absorbing disk lying on the absorbing base, one simply deals with a
whole flat absorbing surface at $z = 0$.  As a consequence, the only
nontrivial behavior emerges for small $\rho$ and large $h$.  Moreover,
even in this region, the offset parameter $\zeta_\infty(\rho,h)$
remains small as compared to $L_2/R_2 \gg 1$, yielding only a moderate
decrease of the total flux.  This is in sharp contrast to the behavior
in the case of the reflecting base shown in
Fig. \ref{fig:Jsurf_kappa0}.  We also stress that the relative error
of the approximate flux from Eq. (\ref{eq:Japp}) is below $1\%$ for
the whole considered range of parameters $\rho$ and $h$.  In other
words, the diagonal approximation is remarkably accurate in the case
of the absorbing base.

\subsubsection*{Thin pillar approximation}

When the pillar is thin ($\rho \ll 1$), one can use again
Eq. (\ref{eq:zeta_rho0}), which reads for the absorbing base as
\begin{equation}
\zeta_\infty(\rho,h) \approx \frac{\tanh(\alpha_{0,1}h)}{\alpha_{0,1}} \,.  
\end{equation}
One can distinguish two cases: \\ 
(i) If $h = \infty$, one has 
\begin{equation}
\zeta_\infty(\rho,h) \approx \frac{1}{\alpha_{0,1}} \approx \frac{\sqrt{\ln(1/\rho) - 3/4}}{\sqrt{2}} \,, 
\end{equation}
which slowly diverges as $\rho \to 0$.  Expectedly, as the base is
infinitely far away, its reactivity does not matter, and we retrieve
the behavior given by Eq. (\ref{eq:zeta_rho0_kappa0_hinf}) for the
reflecting base.  \\
(ii) If $h < \infty$, the expansion of the hyperbolic tangent yields
$\zeta_\infty(\rho,h) \approx h$, i.e., there is no divergence, and
the offset reaches a finite limit equal to $h$.  In fact, even if the
pillar has vanished, the particle can react on the base, and the
offset increases (i.e., the total flux decreases) when the base is
getting further from the source, as expected.  Note that the same
behavior also holds for a partially reactive base ($0 < \kappa <
\infty$).

\subsubsection*{Diffusional screening}

\begin{figure}
\begin{center}
\includegraphics[width=85mm]{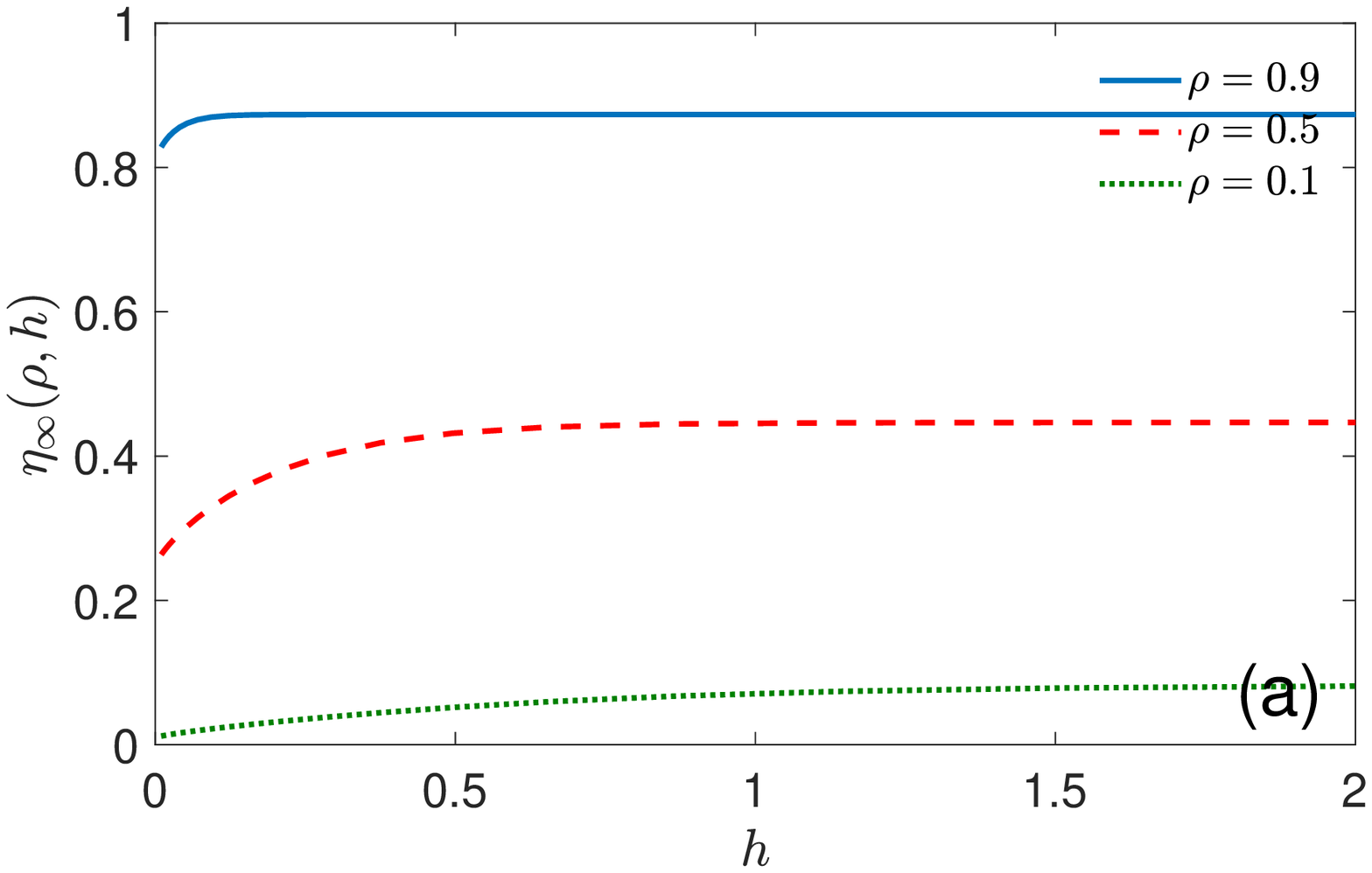} 
\includegraphics[width=85mm]{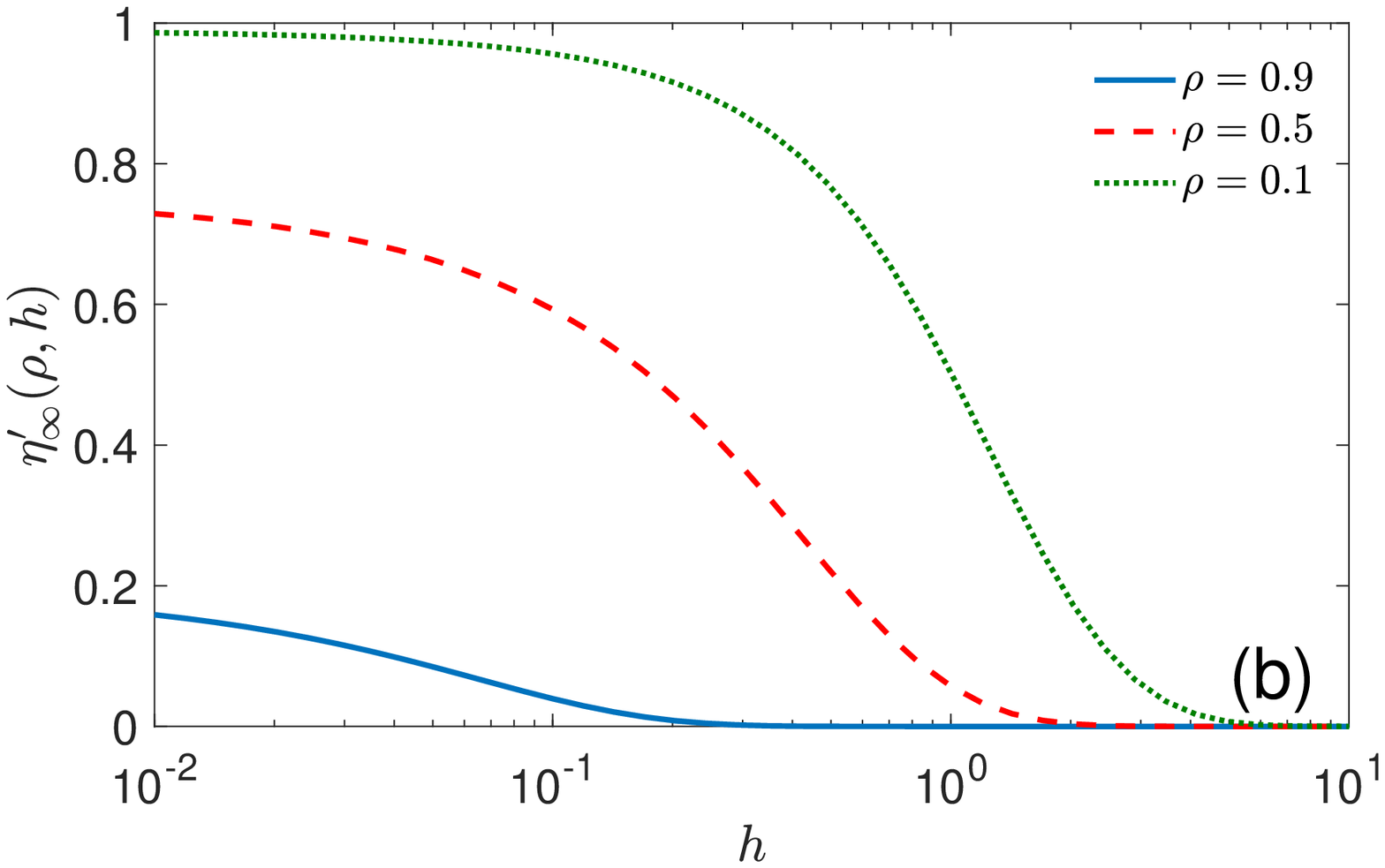} 
\end{center}
\caption{
{\bf (a)} The fraction $\eta_\infty(\rho,h)$ of particles absorbed on
the top of the pillar on the absorbing base ($\kappa = \infty$), as a
function of the pillar height $h = L_1/R_1$, for three values of $\rho
= R_1/R_2$ (see the legend).  This fraction was obtained from
Eq. (\ref{eq:eta}) with $L_2/R_2 = 5$ and $N = 50$.  {\bf (b)} The
fraction $\eta'_\infty(\rho,h)$ of particles absorbed on the absorbing
base as a function of $h$.  This fraction was obtained from
Eq. (\ref{eq:Jbase}) with $L_2/R_2 = 5$ and $N = 50$.}
\label{fig:Jtop_kappainf}
\end{figure}

As the top of the pillar is the most exposed to the source, it may
absorb a considerable fraction of particles and thus screen the
cylindrical part of the pillar and the base.  Panel (a) of
Fig. \ref{fig:Jtop_kappainf} presents the fraction
$\eta_\infty(\rho,h)$ of particles absorbed on the top of the pillar.
When $h = 0$, the disk-like pillar lies on the absorbing base and
therefore absorbs the fraction of particles that is given by its
relative surface area, i.e., $\eta_\infty(\rho,0) = \rho^2$.  As the
pillar's height increases, one might expect that this fraction would
diminish, as the total area of the absorbing surface increases due to
the emergence of the cylindrical part of the pillar.  However, one
sees the opposite trend, i.e., $\eta_\infty(\rho,h)$ actually
increases with $h$.  We already discussed in Sec. \ref{sec:general}
that the cylindrical surface of the pillar, as well as the absorbing
base itself, are getting less accessible due to diffusional screening
when $h$ increases.  Panel (a) re-confirms this behavior by showing a
monotonous growth of $\eta_\infty(\rho,h)$ with $h$.  In the limit
$h\to \infty$, $\eta_\infty(\rho,h)$ approaches a finite value, which
is independent of the base reactivity $\kappa$.  Its dependence on
$\rho$ was illustrated on Fig. \ref{fig:Jtop}(b).

As the base is absorbing, it is instructive to quantify the fraction
of particles absorbed on that base, i.e., $J_{\rm base}/J$.  The limit
$\eta'_\infty(\rho,h)$ of this fraction as $L_2\to\infty$ is shown on
panel (b) of Fig. \ref{fig:Jtop_kappainf}.  At $h = 0$, one simply has
$\eta'_\infty(\rho,0) = 1 - \rho^2$.  When $h$ increases, the base is
getting further from the source so that this fraction should decrease.
According to our exact expression (\ref{eq:Jbase}), this fraction
vanishes exponentially fast, i.e., $\eta'_\infty(\rho,h) \propto
e^{-\alpha_{0,1} h}$ as $h\to \infty$.  In turn, the rate of this
exponential decay, given by $\alpha_{0,1}$ from Eq. (\ref{eq:alpha1}),
slowly vanishes as $\rho \to 0$ according to Eq. (\ref{eq:alpha01}).
In fact, as the pillar gets thinner, it has a lower capacity to
capture particles, which have therefore higher chances to reach a
remote absorbing base.

\subsubsection*{Comparison to hemispheroidal bosses}

It is instructive to compare our exact solution to perturbative
results by Sarkar and Prosperetti \cite{Sarkar_1995}, who considered
the square lattice of hemispheroidal ``bosses'' of height $L_1$ and
radius $R_1$ sitting on the flat absorbing base.  Among many
quantities, they calculated the position of an equivalent flat
boundary, which reads in our notations as
\begin{equation}
\label{AB5}
z_0  = L_1 - (1 + k) v ,
\end{equation}
where $v$ is the volume of the ``bosses'' per unit area occupied by
them, $v = \frac{2\pi}{3} R_1^2 L_1/(\pi R_2^2)$, and
\begin{equation}
\label{AB6}
k  =  - \frac{\Lambda^{-2} \arctan (\sqrt{\Lambda^{-2} - 1})  - \sqrt{\Lambda^{-2} - 1} }{\Lambda^{-2} [\arctan (\sqrt{\Lambda^{-2} - 1})
- \sqrt{\Lambda^{-2} - 1}]} \,,
\end{equation}
with $\Lambda = L_1/R_1 = h/\rho$ being the aspect ratio of the
``boss'', which can be smaller than $1$ for oblate spheroids, and
larger than $1$ for prolate spheroids.  In the limit $h \ll \rho$
(i.e., $\Lambda \to 0$), one gets $k \approx
\Lambda(\pi/2) + O(\Lambda^2)$ so that 
\begin{equation*}
\frac{z_0}{R_2} \approx h - \frac{(1+k)v}{R_2} \approx h - \frac{v}{R_2} = (1-(2/3) \rho^2)h \to 0, 
\end{equation*}
and one retrieves zero offset for a flat absorbing surface, as
expected.

In the opposite limit $h\gg \rho$ (i.e., $\Lambda \gg 1$), one gets $k
\approx \Lambda^2/(\ln(2\Lambda) - 1)$ so that
\begin{equation}
\label{AB63}
\zeta_\infty(\rho,h) = \frac{z_0}{R_2} \approx h - \frac{2}{3} \, \frac{h^3}{\ln(2h/\rho)-1} \,.
\end{equation}
When $h$ is large enough, this expression is negative that indicates
the failure of this approximation for elongated bosses.  This is not
surprising given the perturbative character of this approximation.
This failure highlights the challenges in getting an accurate
theoretical description of diffusive flux towards a spiky coating in
three dimensions that we managed to resolve in this paper.

\section{Conclusion}
\label{sec:conclusion}

We studied steady-state diffusion towards a periodic array of
absorbing pillars protruding from a flat base.  We developed an
analytical framework for getting an equivalent flat absorbing boundary
that maintains the same diffusive flux through the system as the
original spiky boundary when the flux is generated by a remote source.
The proposed analytical solution accounts for the adsorbing spikes and
the absorbing or reflecting base of the coating as well as for the
more general case of partially reactive base.  For this purpose, we
approximated the solution in the original setting with a periodic
arrangement of spikes by that to a simpler problem with a single
absorbing cylindrical pillar inside a reflecting cylindrical tube.
The reduced problem was then solved exactly by a mode matching method.
In this way, we obtained an exact solution $\c(r,z)$, in which the
dependence on the spatial point $(r,z)$ is captured via explicit
analytical functions, while the coefficients in front of these
functions have to be obtained by a numerical inversion of a matrix
with explicitly known elements.  Despite this numerical step, the
exact solution allows one to investigate the trapping efficiency of
the spiky coating in different asymptotic regimes.  Moreover, we
proposed a diagonal approximation, which eliminates the numerical
matrix inversion and thus yields an analytical solution.  We showed
that this approximation is very accurate for a broad range of
parameters.

Our main focus was on the dependence of the total flux of particles on
the geometric parameters such as the pillar's height $L_1$ and radius
$R_1$, the inter-pillar distance $\ell$ (or $R_2$), and the distance
to the source $L_2$.  We showed that the total flux is given by a
simple expression (\ref{eq:Jexact}), in which the whole geometric
complexity is captured via the offset parameter $z_0$.  This parameter
does not almost depend on $L_2$ whenever $L_2\gg R_2$.  We then
analyzed the dependence of $z_0$ on the rescaled height $h$ and radius
$\rho$ of the pillar.  For instance, we obtained a logarithmically
slow increase of the offset $z_0$ in the thin pillar limit ($\rho \to
0$).  This is in contrast to a much faster increase, as $1/\rho$, in
the case of a disk-like pillar (with $h = 0$).  This observation
highlights the important role of the absorber anisotropy.  We also
discussed some earlier approaches such as the capacitance
approximation or a perturbative analysis for hemispheroidal spikes.
In addition to the total flux, we also considered some refined
characteristics such as the flux on the top of the pillar and the flux
on the absorbing base.  These fluxes allowed us to quantity the
diffusional screening when the most exposed parts of the absorbing
surface (such as the top of the pillar) capture a considerable
fraction of particles and therefore screen less exposed parts, in a
direct analogy with electrostatic screening.

While the present work was concentrated on the steady-state diffusion
towards a spiky coating, the developed method can be generalized in
several ways.  In Appendix \ref{sec:extensions}, we briefly described
four immediate extensions of the present work: (i) replacement of the
Laplace equation $\Delta u = 0$ by the modified Helmholtz equation
$(\Delta - p/D) u = 0$ that allows one to incorporate bulk reactivity
$p$ or, equivalently, a finite lifetime of diffusing particles
\cite{Yuste13,Meerson15,Grebenkov17d}; moreover, one can also study
nonstationary diffusion with the help of an inverse Laplace transform
of the solution with respect to $p$; this framework may also be
valuable for developing acoustic materials covered a soft brush
(rubber spikes); (ii) replacement of the uniform source $u(r,L_2) = 1$
by an arbitrary source distribution; (iii) replacement of the
Dirichlet boundary condition $u(r,L_2) = 1$ by Neumann boundary
condition $(\partial_z u(r,z))_{z=L_2} = 0$ that makes the top disk
reflecting; together with the first extension, it gets access to the
probability density function of the first-passage time on the pillar.
A separate work will be devoted to studying this quantity
\cite{Grebenkov_2023}.  Finally, (iv) the employed mode matching
method can be used to treat a cylindrical pillar of arbitrary
cross-section inside an outer reflecting tube of arbitrary
cross-section; in this case, the explicit radial functions have to be
replaced by appropriate Laplacian eigenfunctions in cross-sections.

On the application side, the presented analytical results can be a
useful tool for building simple mathematical models of diffusive
transport near rough interfaces and tailored design of new
meta-material coatings.  It is worth noting that similar problems
emerge in many other areas of the Laplacian transport such as
electrostatics, fluid dynamics, and heat transfer
\cite{Bernard_2001,Vandembroucq_1997,Bazant_2016,Blyth_1993,Blyth_2003,Fyrillas_2001,Marigo_2016}.  
The results of the present study can easily be translated to this
broader context.

While we gave a systematic analysis of the trapping properties of a
spiky coating, many aspects of this challenging problem need further
investigations.  On the mathematical side, it would be important to
derive rigorously some asymptotic regimes (e.g., the capacitance
approximation) from our exact solution.  This analysis requires
refined mathematical tools to deal with spectral sums involving zeros
of Bessel functions \cite{Grebenkov19}.  From the applicative
standpoint, one can inspect the case of a partially reactive base,
which lies in between the considered cases of reflecting and absorbing
bases.  The reactivity of the base adds a tunable parameter, which may
adjust and control the trapping efficiency of the spiky coating.  In
addition, one can adapt the present description to treat other
arrangements of the pillars (e.g., a hexagonal lattice).  Finally, as
heterogeneities in pillar's properties may considerably affect the
total flux, their analysis presents an interesting perspective of the
present work.

\begin{acknowledgments}
D.S.G. acknowledges the Alexander von Humboldt Foundation for support
within a Bessel Prize award.  A.T.S. thanks Paul A. Martin for many
illuminating discussions.
\end{acknowledgments}

%


\appendix

\section{Exact solution}
\label{sec:exact}

In this Appendix, we provide the details of derivation of the exact
solution of the boundary value problem (\ref{eq:problem_u}).

\subsection{Derivation of the solution}

Due to the axial symmetry, the boundary value problem
(\ref{eq:problem_u}) is actually a two-dimensional problem in an
L-shape region (see Fig. \ref{fig:scheme}(c)).  Note that one has to
add the Neumann boundary condition,
\begin{equation}  \label{eq:cond_r0}
(\partial_r u(r,z))_{r = 0} = 0 \quad (0 < z < L_2),
\end{equation} 
to account for the regularity and axial symmetry of the problem.  One
can search for its solution separately in two rectangular subdomains,
$\Omega_1 = (R_1,R_2) \times (-L_1,0)$ and $\Omega_2 = (0,R_2) \times
(0,L_2)$, and then match them at the junction interval at $z = 0$.

A general solution in $\Omega_1$ reads
\begin{equation}   \label{eq:u1}
u_1(r,z) = \sum\limits_{n=0}^\infty c_{n,1} \, v_{n,1}(r/R_2) \, \frac{s_{n,1}(z)}{s_{n,1}(0)} \,,
\end{equation}
with unknown coefficients $c_{n,1}$, where
\begin{align}  \nonumber
s_{n,1}(z) & = \sinh\biggl(\frac{\alpha_{n,1} (L_1+z)}{R_2}\biggr)  \\  \label{eq:s1}
& + \frac{\alpha_{n,1}}{\kappa} \cosh\biggl(\frac{\alpha_{n,1} (L_1+z)}{R_2}\biggr)
\end{align}
and
\begin{equation}
v_{n,1}(\r) = e_n \, w_n(\r) ,
\end{equation}
with
\begin{equation}
w_n(\r) = J_1(\alpha_{n,1}) Y_0(\alpha_{n,1} \r) - Y_1(\alpha_{n,1}) J_0(\alpha_{n,1} \r),
\end{equation}
and we used $J'_0(z) = -J_1(z)$, $Y'_0(z) = -Y_1(z)$, prime denotes
the derivative, $\r$ denotes dimensionless radius, $\kappa = R_2
\kappa_b/D$ is the dimensionless reactivity of the base, and $J_\nu(z)$
and $Y_\nu(z)$ are the Bessel functions of the first and second kind,
respectively.  The prefactor
\begin{equation}
e_n = \frac{\sqrt{2}}{\sqrt{[w_n(1)]^2 - \rho^2 [w'_n(\rho)/\alpha_{n,1}]^2}}
\end{equation}
ensures the normalization:
\begin{equation}
\int\limits_\rho^1 d\r \, \r \, [v_{n,1}(\r)]^2 = 1 ,
\end{equation}
where $\rho = R_1/R_2$.  Here we used
\begin{align*}
\int\limits_\rho^1 d\r \, \r\, w_n^2(\r) & = \frac{1}{2\alpha_{n,1}^2} 
\biggl(\r^2 [w'_n(\r)]^2 + \alpha_{n,1}^2 \r^2 [w_n(\r)]^2\biggr)_\rho^1 \\ & = \frac{[w_n(1)]^2 - \rho^2 [w'_n(\rho)/\alpha_{n,1}]^2}{2} ,
\end{align*}  
with $w_n(\rho) = 0$ and $w'_n(1) = 0$ being employed.
By construction, $u_1(r,z)$ is a harmonic function that satisfies
Eqs. (\ref{eq:cond_outer}, \ref{eq:cond_bottom}).  The parameters
$\alpha_{n,1}$ are obtained by imposing the condition
(\ref{eq:cond_inner}) at $r = R_1$ (i.e., setting $w_n(\rho) = 0$) and
solving the resulting equation
\begin{equation}  \label{eq:alpha1}
Y_1(\alpha_{n,1}) J_0(\alpha_{n,1} \rho) - J_1(\alpha_{n,1}) Y_0(\alpha_{n,1} \rho) = 0  .
\end{equation}
This equation has infinitely many positive solutions
$\{\alpha_{n,1}\}$, which are enumerated by $n=0,1,2,\ldots$ in an
increasing order \cite{Watson}.  As $v_{n,1}(\r)$ are the
eigenfunctions of the differential operator $\partial_r^2 + (1/r)
\partial_r$, they form a complete orthonormal basis in the space
$L_2(\rho,1)$.

A general solution in $\Omega_2$ reads
\begin{equation}  \label{eq:u2}
u_2(r,z) = 1 - \sum\limits_{n=0}^\infty c_{n,2}\, v_{n,2}(r/R_2) \, s_{n,2}(z) ,
\end{equation}
with unknown coefficients $c_{n,2}$, where
\begin{equation}  \label{eq:v2}
v_{n,2}(\r) =  \frac{J_0(\alpha_{n,2}\r)}{J_0(\alpha_{n,2})} \, ,
\end{equation} 
and
\begin{equation} \label{eq:s2}
s_{n,2}(z) = \frac{\sinh(\alpha_{n,2} (L_2-z)/R_2)}{\sinh(\alpha_{n,2} L_2/R_2)} \,. 
\end{equation}
By construction, $u_2(r,z)$ is a harmonic function that satisfies
Eqs. (\ref{eq:cond_top}, \ref{eq:cond_r0}).  The parameters
$\alpha_{n,2}$ are obtained by imposing the condition
(\ref{eq:cond_outer}):
\begin{equation}  \label{eq:alpha_n2}
J_1(\alpha_{n,2}) = 0  \quad (n=0,1,2,\ldots).
\end{equation}
This equation has infinitely many positive solutions $\{
\alpha_{n,2}\}$, which are enumerated by $n = 0,1,2,\ldots$ in an
increasing order \cite{Watson}.  Note that $\alpha_{0,2} = 0$ and the
corresponding term in Eq. (\ref{eq:u2}) is $c_{0,2} (L_2-z)/L_2$.  The
prefactor in Eq. (\ref{eq:v2}) ensures the normalization:
\begin{equation}  \label{eq:v2norm}
\int\limits_0^1 d\r \, \r \, [v_{n,2}(\r)]^2 = \frac{1}{2} \,.
\end{equation}
As $\sqrt{2} \, v_{n,2}(\r)$ are the eigenfunctions of the
differential operator $\partial_r^2 + (1/r) \partial_r$, they form a
complete orthonormal basis in the space $L_2(0,1)$.

The unknown coefficients $c_{n,1}$ and $c_{n,2}$ are then determined
by matching these solutions at $z = 0$, i.e., by requiring the
continuity of $u$ and of its derivative $\partial_z u$.  The second
condition, which should be satisfied for any $R_1 < r < R_2$, reads
\begin{align}  \nonumber
& R_2 (\partial_z u_1)_{z=0} = \sum\limits_{n=0}^\infty \tilde{c}_{n,1} v_{n,1}(r/R_2) \,  \\  \label{eq:matching2}
& = \sum\limits_{n=0}^\infty \tilde{c}_{n,2} v_{n,2}(r/R_2) = R_2 (\partial_z u_2)_{z=0} ,
\end{align}
where
\begin{align*}
\tilde{c}_{n,1} & = c_{n,1} R_2 \frac{s'_{n,1}(0)}{s_{n,1}(0)} \,, \\
\tilde{c}_{n,2} & = c_{n,2} B_n^{(2)} ,  
\end{align*}
with $h = L_1/R_2$,
\begin{align*}
s'_{n,1}(0) & = \frac{\alpha_{n,1}}{R_2} \biggl[\cosh(\alpha_{n,1} h) 
 + \frac{\alpha_{n,1}}{q R_2} \sinh(\alpha_{n,1} h)\biggr],
\end{align*}
and 
\begin{subequations}
\begin{align}  \label{eq:Bn2}
B_n^{(2)} & = - R_2 s'_{n,2}(0) = \alpha_{n,2}\, \ctanh(\alpha_{n,2} L_2/R_2) , \\   \label{eq:B20}
B_0^{(2)} &= R_2/L_2  .
\end{align}
\end{subequations}
Multiplying Eq. (\ref{eq:matching2}) by $\r\, v_{k,1}(\r)$ and
integrating from $\rho$ to $1$, one gets
\begin{align*}
& \sum\limits_{n=0}^\infty \tilde{c}_{n,2} \int\limits_\beta^1 d\r \, \r \, v_{k,1}(\r) \,  v_{n,2}(\r) 
 = \tilde{c}_{k,1} 
\end{align*}
due to orthogonality of $\{v_{k,1}(\r)\}$.  Setting
\begin{equation}   \label{eq:A_def}
A_{k,n} = \int\limits_\rho^1 d\r \, \r \, v_{k,1}(\r) \,  v_{n,2}(\r) , 
\end{equation}
we can rewrite the above equations as
\begin{equation}  \label{eq:ck1}
c_{k,1} = B_k^{(1)} \sum\limits_{n=0}^\infty A_{k,n} B_n^{(2)} c_{n,2} \,,
\end{equation} 
where 
\begin{equation}   \label{eq:Bn1}
B_n^{(1)} = \frac{s_{n,1}(0)}{R_2 s'_{n,1}(0)} = \frac{1}{\alpha_{n,1}} \, \frac{\alpha_{n,1} + \kappa \tanh(\alpha_{n,1} h)}
{\kappa + \alpha_{n,1} \tanh(\alpha_{n,1} h)}  \,.
\end{equation}
Moreover, as the radial functions $v_{k,1}(\r)$ and $v_{k,2}(\r)$ are
linear combinations of Bessel functions of the {\it same order}, the
integral in Eq. (\ref{eq:A_def}) can be found explicitly:
\begin{align}  \nonumber
A_{k,n} & = \biggl(\r\frac{v_{k,1}(\r) v'_{n,2}(\r) - v'_{k,1}(\r) v_{n,2}(\r)}{\alpha_{k,1}^2 - \alpha_{n,2}^2}\biggr)_{\r = \rho}^{1} \\  \label{eq:A}
& = \frac{\rho\, v'_{k,1}(\rho)\, v_{n,2}(\rho)}{\alpha_{k,1}^2 - \alpha_{n,2}^2} \,,
\end{align}
where we used the boundary conditions $v_{k,1}(\rho) = v'_{k,1}(1) =
v'_{n,2}(1) = 0$.

\begin{figure}
\begin{center}
\includegraphics[width=30mm]{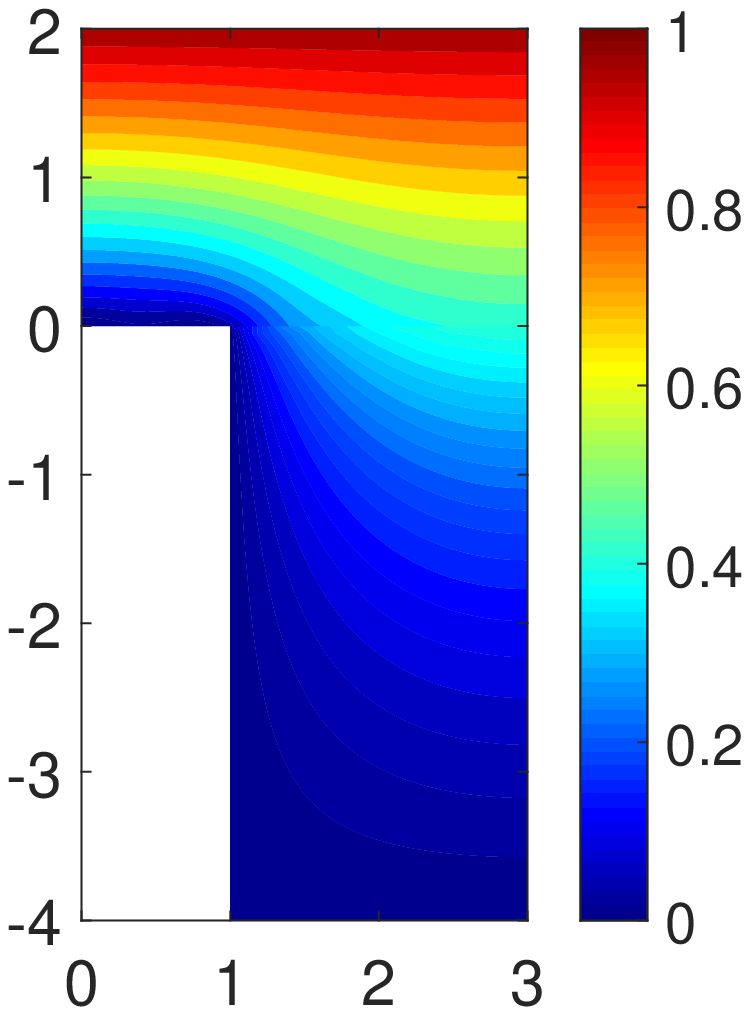} 
\hskip 5mm
\includegraphics[width=30mm]{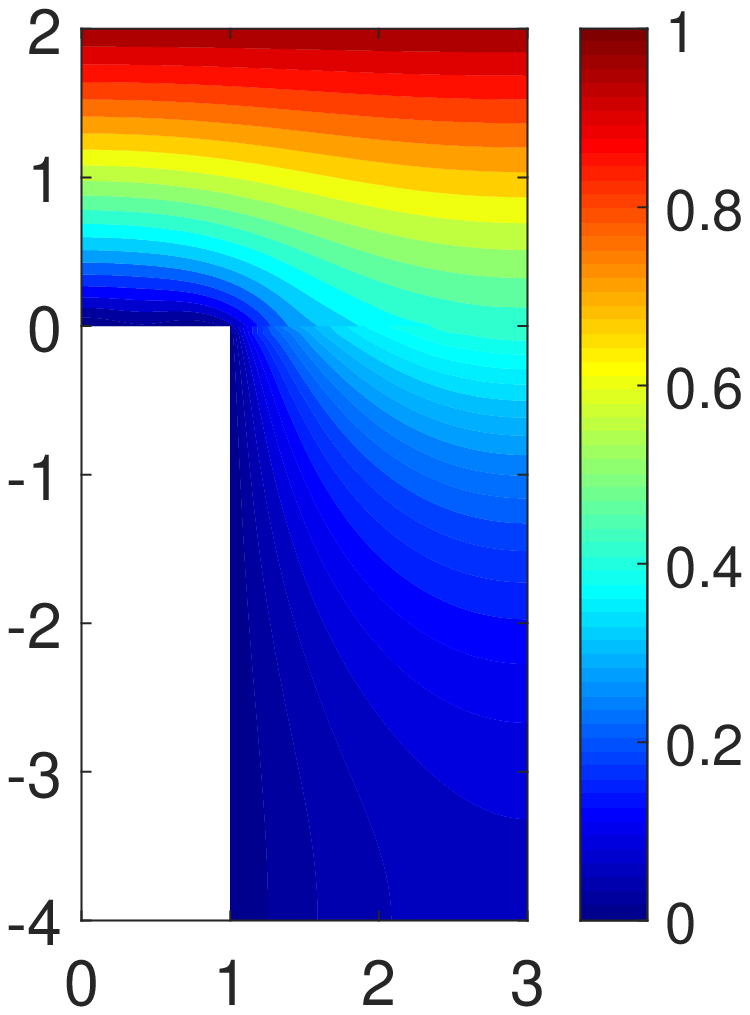} 
\end{center}
\caption{
Planar ($xz$) projection of the solution $u(r,z)$ for $L_1 = 4$, $R_1
= 1$, $L_2 = 2$, $R_2 = 3$, and two values of the base reactivity:
$\kappa = \infty$ {\bf (left)} and $\kappa = 0$ {\bf (right)}.  We
used the truncation order $N = 10$.}
\label{fig:examples1}
\end{figure}

\begin{figure}
\begin{center}
\includegraphics[width=80mm]{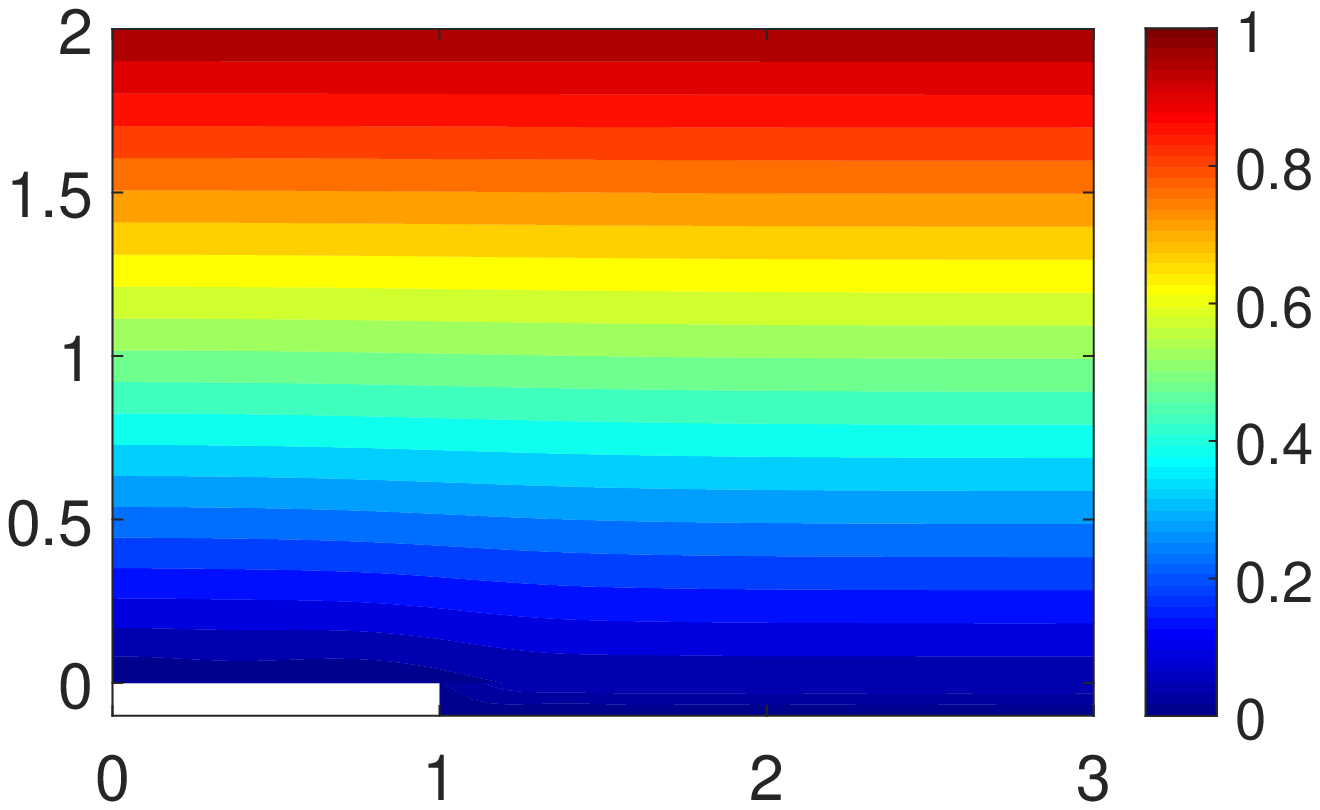} 
\includegraphics[width=80mm]{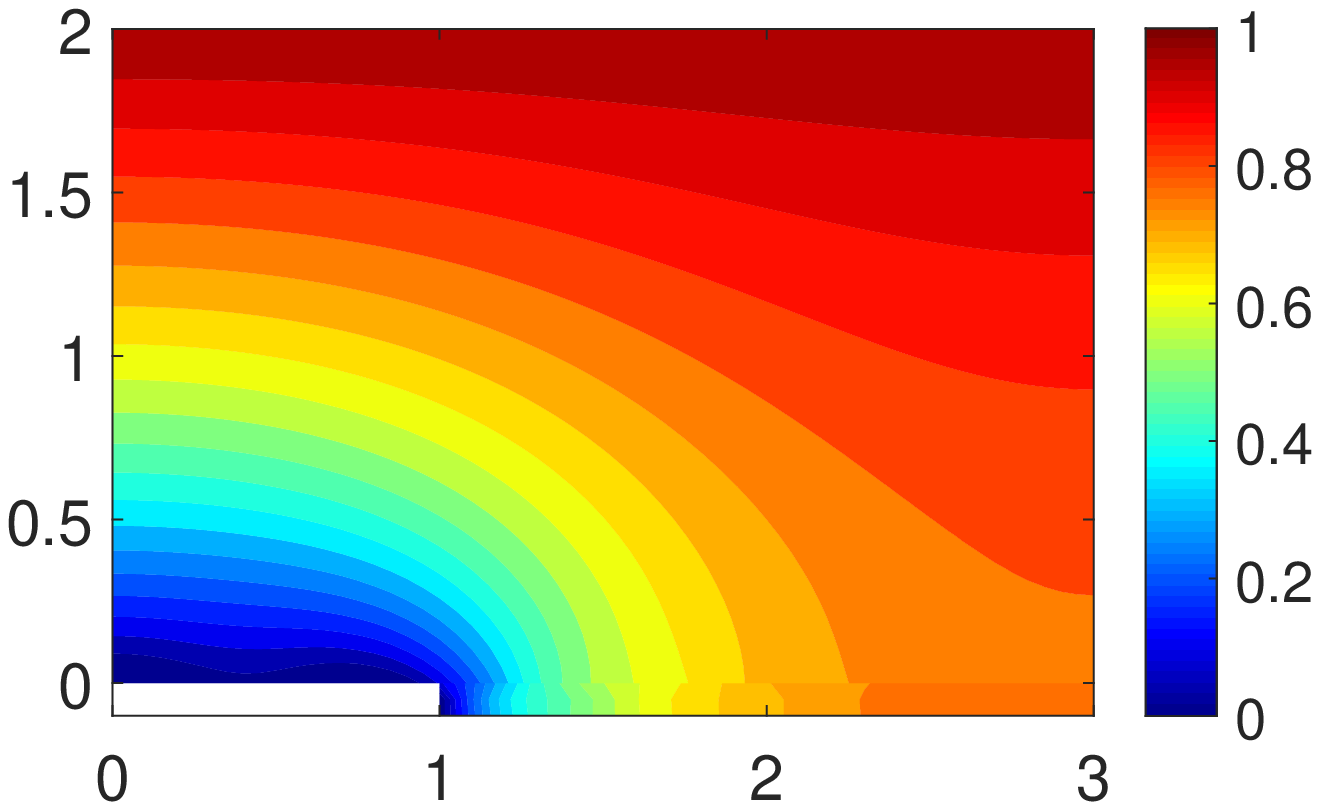} 
\end{center}
\caption{
Planar ($xz$) projection of the solution $u(r,z)$ for $L_1 = 0.1$,
$R_1 = 1$, $L_2 = 2$, $R_2 = 3$, and two values of the base
reactivity: $\kappa = \infty$ {\bf (top)} and $\kappa = 0$ {\bf
(bottom)}.  We used the truncation order $N = 10$.}
\label{fig:examples2}
\end{figure}

Similarly, we impose the continuity of the function $u(r,z)$ at $z=0$,
together with Eq. (\ref{eq:cond_disk}):
\begin{equation}  \label{eq:u_continuity}
u_2(r,0) = \begin{cases} 0  \hskip 19mm (0 < r < R_1), \cr   u_1(r,0) \qquad (R_1 < r < R_2). \end{cases}
\end{equation}
Multiplying this relation by $\r v_{k,2}(\r)$ and integrating from
$0$ to $1$, we get
\begin{align*}
& \int\limits_0^1 d\r \, \r \, v_{k,2}(\r) - \frac{c_{k,2}}{2}
 = \int\limits_{\rho}^1 d\r \, \r \, v_{k,2}(\r) \sum\limits_{n=0}^\infty c_{n,1} \, v_{n,1}(\r),
\end{align*}
where we used the orthogonality of functions $\{v_{k,2}(\r)\}$ and
their normalization (\ref{eq:v2norm}).  Since $v_{0,2}(\r) = 1$, the
orthogonality of functions $\{v_{k,2}(\r)\}$ implies that the first
integral vanishes for all $k > 0$, and yields $1/2$ for $k = 0$:
\begin{equation}
\frac{c_{k,2}}{2} + \sum\limits_{n=0}^\infty  c_{n,1} A_{n,k} = \frac{\delta_{k,0}}{2} \qquad (k = 0,1,2,\ldots).
\end{equation}
Substituting $c_{n,1}$ from Eq. (\ref{eq:ck1}), we get
\begin{equation*} 
c_{k,2} + 2\sum\limits_{n=0}^\infty A_{n,k} B_n^{(1)} \sum\limits_{n'=0}^\infty A_{n,n'} B_{n'}^{(2)} c_{n',2} = \delta_{k,0} \,.
\end{equation*}
It is convenient to re-arrange two sums as
\begin{equation} \label{eq:system}
c_{k,2} + \sum\limits_{n=0}^\infty W_{k,n}\, c_{n,2}  = \delta_{k,0}  \quad (k = 0,1,2,\ldots),
\end{equation}
where
\begin{equation}  \label{eq:Wdef}
W_{k,n} = 2\sum\limits_{n'=0}^\infty A_{n',k} B_{n'}^{(1)}  A_{n',n} B_{n}^{(2)} \,,
\end{equation}
i.e., we got the infinite system of linear algebraic equations for the
unknown coefficients $c_{k,2}$ with $k=0,1,2,\ldots$.  To compute
these coefficients, one needs to construct the infinite-dimensional
matrix $W$ and then to invert the matrix $I + W$, where $I$ is the
identity matrix.  In practice, one can truncate the matrix $W$ to a
finite size $N\times N$ and then perform the inversion of $I+W$
numerically.  Once the coefficients $c_{n,2}$ are found, one can
determine $c_{n,1}$ according to Eq. (\ref{eq:ck1}).  This completes
the construction of the exact solution of the problem
(\ref{eq:problem_u}).  Even though this construction involves
numerical inversion of the truncated matrix, the obtained expressions
(\ref{eq:u1}, \ref{eq:u2}) provides an explicit analytical dependence
of $u(r,z)$ on $r$ and $z$ via the functions $v_n(r/R_2)$ and
$s_n(z)$.  Moreover, the accuracy of the numerical construction of
$u(r,z)$ rapidly improves as the truncation order $N$ increases.  In
most cases, one can use moderate values of $N$ (say, few tens) to get
very accurate results.  Figures \ref{fig:examples1} and
\ref{fig:examples2} illustrate the exact solution for two
configurations with long and short pillars.  In the first case, the
source was placed very close to the pillar to highlight changes in the
concentration profile.

It is also instructive to consider the concentration averaged over the
cross-section (denoted by bar):
\begin{equation}  \label{eq:u2_bar}
\overline{u}_2(z) = \frac{2\pi}{\pi R_2^2} \int\limits_0^{R_2} dr \, r \, u_2(r,z) = 1 - c_{0,2}  \, \frac{L_2-z}{L_2}
\end{equation}
and
\begin{align}  \nonumber
\overline{u}_1(z) & = \frac{2\pi}{\pi (R_2^2-R_1^2)} \int\limits_{R_1}^{R_2} dr \, r \, u_1(r,z) \\  \nonumber
& = \frac{2}{1-\rho^2} \sum\limits_{n=0}^\infty c_{n,1} \frac{s_{n,1}(z)}{s_{n,1}(0)} \int\limits_{\rho}^1 d\r \, \r \, v_{n,1}(\r) \\
& = \frac{2}{1-\rho^2} \sum\limits_{n=0}^\infty c_{n,1} \frac{s_{n,1}(z)}{s_{n,1}(0)} \frac{\rho v'_{n,1}(\rho)}{\alpha_{n,1}^2} \,,
\end{align}
where we used
\begin{equation}
\int\limits_{\rho}^1 d\r \, \r\, v_{n,1}(\r) = \frac{\rho\, v'_{n,1}(\rho)}{\alpha_{n,1}^2}
\end{equation}
due to the boundary conditions.  While the averaged solution in the
upper part is indeed linear, this is not true for the lower part.

\subsection{Total flux}
\label{sec:distant}

The total flux can be found as
\begin{align} \nonumber
J & = 2\pi \c_0 \int\limits_0^{R_2} dr \, r\, (D \partial_z u_2)_{z=L_2} \\  \nonumber
& = 2\pi D\c_0  \int\limits_0^{R_2} dr \, r\, \sum\limits_{n=0}^\infty c_{n,2} v_{n,2}(r/R_2) \frac{B_n^{(2)}}{R_2} \\
\label{eq:J}
& = \frac{\pi D \c_0 R_2^2}{L_2} \, c_{0,2} \,,
\end{align}
where $\c_0$ is the imposed concentration of particles at the top, and
we used the orthogonality of functions $\{v_{n,2}(\r)\}$.  One sees
that the coefficient $c_{0,2}$ incorporates the sophisticated
dependence of $J$ on the geometric parameters $L_1$, $R_1$, $L_2$, and
$R_2$.

The structure of the exact solution reveals how different geometric
parameters can affect the concentration and the total flux.  In
particular, the pillar height $L_1$ enters {\it only} via $B_n^{(1)}$,
the distance to the source $L_2$ enters {\it only} via $B_n^{(2)}$, so
that the matrix $A$ does not depend on $L_1$ and $L_2$.  In the limit
$L_2\to\infty$, one gets $B_n^{(2)} \to \alpha_{n,2}$ for all $n > 0$.
In fact, Eq. (\ref{eq:Bn2}) implies that if $L_2 \gg R_2$,
$\ctanh(\alpha_{n,2} L_2/R_2)$ for $n > 0$ can be replaced by $1$, and
the error of this replacement is exponentially small (note that
$\alpha_{n,2} \geq \alpha_{1,2} \approx 3.8317$ for $n \geq 1$).  In
other words, the elements $B_n^{(2)}$ does not depend on $L_2$
whenever $L_2\gg R_2$, except for $B_0^{(2)} = R_2/L_2$.

This property suggests to treat separately the equations for $n = 0$
and $n > 0$ in the system (\ref{eq:system}).  Let us denote by $W'$
the reduced matrix obtained by removing from $W$ the column and the
row corresponding to $n=0$.  More explicitly, we rewrite
Eq. (\ref{eq:system}) as
\begin{align*}
c_{0,2} + W_{0,0} c_{0,2} + \sum\limits_{n=1}^\infty W_{0,n} c_{n,2} & = 1 , \\
c_{k,2} + \sum\limits_{n=1}^\infty W_{k,n} c_{n,2} & = - W_{k,0} c_{0,2}  \quad (k = 1,\ldots),
\end{align*}
solve the second set of equations by inverting the matrix $I+W'$,
substitute the solution in the first line and then express $c_{0,2}$
as
\begin{equation}
c_{0,2} = \biggl(1 + W_{0,0} - \sum\limits_{k,k'=1}^\infty W_{0,k} \bigl[(I+W')^{-1}\bigr]_{k,k'} W_{k',0} \biggr)^{-1} .
\end{equation}
Importantly, both terms $W_{0,0}$ and $W_{k',0}$ are proportional to
$R_2/L_2$ due to Eq. (\ref{eq:B20}).  We can therefore represent this
coefficient as
\begin{equation}  \label{eq:c02_bis}
c_{0,2} = \frac{1}{1 + z_0/L_2} \,,
\end{equation}
where
\begin{equation}  \label{eq:z0}
z_0 = L_2\biggl(W_{0,0} - \sum\limits_{k,k'=1}^\infty W_{0,k} \bigl[(I+W')^{-1}\bigr]_{k,k'} W_{k',0} \biggr).
\end{equation}
While the representation (\ref{eq:c02_bis}) is valid for any $L_2$,
its main advantage is that $z_0$ rapidly approaches a constant as
$L_2\to \infty$, allowing one to eliminate this geometric parameter.

\subsection{Trapping efficiency of the pillar}
\label{sec:Atop}

In order to quantify the trapping efficiency of the pillar, it is
instructive to calculate the flux of particles absorbed on the top of
the pillar:
\begin{align}  \nonumber
J_{\rm top} & = 2\pi \c_0 \int\limits_0^{R_1} dr\, r \, (D \partial_z u)_{z=0} \\  \nonumber
& = 2\pi D\c_0 \int\limits_0^{R_1} dr\, r \sum\limits_{n=0}^\infty c_{n,2} v_{n,2}(r/R_2) \frac{B_n^{(2)}}{R_2} \\
& = 2\pi D\c_0 R_1 \sum\limits_{n=0}^\infty c_{n,2} B_n^{(2)} \frac{J_1(\alpha_{n,2}\rho)}{\alpha_{n,2} J_0(\alpha_{n,2})} \,.
\end{align}
To eliminate the dependence on the distance $L_2$, one can express the
coefficients $c_{n,2}$ in terms of $c_{0,2}$:
\begin{align}  \nonumber
J_{\rm top} & = \pi D \c_0 R_1 c_{0,2} \frac{R_2}{L_2} \biggl(\rho - 2 \sum\limits_{n=1}^\infty 
B_n^{(2)} \frac{J_1(\alpha_{n,2}\rho)}{\alpha_{n,2} J_0(\alpha_{n,2})}  \\
& \times \biggl(\sum\limits_{k=1}^\infty \bigl[(I+W')^{-1}\bigr]_{n,k} \frac{W_{k,0} L_2}{R_2} \biggr) \,,
\end{align}
where we wrote explicitly the term with $n=0$ due to $\alpha_{0,2} =
0$.  Rescaling this expression by the total flux $J$ eliminates
$c_{0,2}$, yielding
\begin{align}  \nonumber
\frac{J_{\rm top}}{J} & = \rho \biggl(\rho - 2\sum\limits_{n,k=1}^\infty  
B_n^{(2)} \frac{J_1(\alpha_{n,2}\rho)}{\alpha_{n,2} J_0(\alpha_{n,2})} \\  \label{eq:Jtop}
& \times \bigl[(I+W')^{-1}\bigr]_{n,k} \frac{W_{k,0} L_2}{R_2}   \biggr).
\end{align}
As $W_{k,0}$ is proportional to $R_2/L_2$, the factor $L_2/R_2$ is
compensated, and we conclude that the ratio $J_{\rm top}/J$ does not
almost depend on $L_2$ whenever $L_2 \gg R_2$.  In other words, the
distance to a remote source does not affect the fraction of particles
absorbed on the top of the pillar that allows us to define
\begin{equation}  \label{eq:eta}
\eta_\kappa(\rho,h) = \lim\limits_{L_2\to\infty} \frac{J_{\rm top}}{J} \,.
\end{equation}

When $L_2/R_2 \gg 1$, one has $B_n^{(2)} \approx \alpha_{n,2}$ and the
factor $B_n^{(2)} \frac{J_1(\alpha_{n,2}\rho)}{\alpha_{n,2}
J_0(\alpha_{n,2})}$ exhibits oscillations.  Even though the series
(\ref{eq:Jtop}) remains well-defined, its convergence can be rather
slow, especially for small $\rho$ and large $h$.  As a consequence,
getting the asymptotic behavior of $\eta_0(\rho,\infty)/\rho$ in the
limit $\rho \to 0$ can be difficult, even numerically (see
Fig. \ref{fig:Jtop}(b) and the related discussion).

Similarly, we can compute the flux on the base:
\begin{align*}
J_{\rm base} & = 2\pi \c_0 \int\limits_{R_1}^{R_2} dr \, r  (D \partial_z u)_{z=-L_1} \\
& = 2\pi D c_0 R_2^2 \sum\limits_{n=0}^\infty c_{n,1} \frac{\alpha_{n,1}}{R_2 s_{n,1}(0)} \int\limits_{\rho}^1 d\r \, \r \, v_{n,1}(\r)  \\
& = 2\pi D c_0 R_1 \sum\limits_{n=0}^\infty c_{n,1} \frac{v'_{n,1}(\rho)}{\alpha_{n,1} s_{n,1}(0)} .
\end{align*}
Expressing $c_{n,1}$ in terms of $c_{n,2}$ and exchanging the order of
summations, one gets
\begin{align*}
J_{\rm base} & = \pi D \c_0 R_2 \sum\limits_{n=0}^\infty c_{n,2}  V_{n} B_n^{(2)} ,
\end{align*}
where
\begin{equation}
V_{n} = 2\rho \sum\limits_{n'=0}^\infty \frac{v'_{n',1}(\rho)}{\alpha_{n',1} s_{n',1}(0)} B_{n'}^{(1)} A_{n',n}  .
\end{equation}
As previously, we separate the term with $n = 0$ from the remaining
contributions, express $c_{n,2}$ in terms of $c_{0,2}$, and divide by
the total flux to get the fraction of particles absorbed on the base:
\begin{equation}  \label{eq:Jbase}
\frac{J_{\rm base}}{J} = V_0 - \sum\limits_{k,n=1}^\infty V_n B_n^{(2)} \bigl[(I+W')^{-1}\bigr]_{n,k}  \frac{W_{k,0} L_2}{R_2} \,.
\end{equation}
Since this expression does not almost depend on $L_2$ whenever $L_2\gg
R_2$, we define the fraction of particles absorbed by the base in the
limit $L_2\to \infty$:
\begin{equation}
\eta'_\kappa(\rho,h) = \lim\limits_{L_2\to\infty} \frac{J_{\rm base}}{J} \,.
\end{equation}

\subsection{Thin pillar asymptotic behavior}
\label{sec:Athin}

In this section, we derive the asymptotic behavior of the total flux
in the limit $R_1 \to 0$.  The radius $R_1$ affects the solution of
the problem in the domain $\Omega_1$, in particular, the solutions
$\{\alpha_{n,1}\}$ of Eq. (\ref{eq:alpha1}) and thus the matrix
elements of $A$ and $B^{(1)}$.

First of all, we stress that the solutions $\{\alpha_{n,1}\}$ determine
the eigenvalues of the two-dimensional Laplace operator in the cross
section of the tube, i.e., in the annulus between the inner absorbing
circle of radius $R_1$ and the outer reflecting circle of radius
$R_2$.  In the limit $R_1 \to 0$, the absorbing circle can be treated
as a ``strong'' perturbation of the disk of radius $R_2$
\cite{Mazya85,Ward93}.  As a consequence, the eigenvalues of the
``perturbed'' problem approach those of the ``unperturbed'' problem,
which are determined by $\alpha_{n,2}$, i.e., $\alpha_{n,1} \to
\alpha_{n,2}$ for all $n$.  As $\alpha_{n,2}$ are strictly positive
for all $n > 0$, one can simply substitute $\alpha_{n,1}$ by
$\alpha_{n,2}$ in the leading-order approximation.  In contrast, the
approach of $\alpha_{0,1}$ to $\alpha_{0,2} = 0$ determines the
asymptotic behavior of the total flux.

Using the asymptotic behavior of the Bessel functions (see p. 974 in
\cite{Gradsteyn}), we expand Eq. (\ref{eq:alpha1}) as
\begin{equation*}
0 = \frac{Y_1(\alpha_{0,1})}{J_1(\alpha_{0,1})} - \frac{2}{\pi}\biggl(\ln (\alpha_{0,1}\rho/2) + \gamma + \ldots \biggr),
\end{equation*}
where $\gamma \approx 0.5772$ is the Euler constant, and we neglected
higher-order terms in $\rho$.  In order to compensate the divergent
term $\ln \rho$, $\alpha_{0,1}$ should vanish.  Expanding the ratio
\begin{equation}
\frac{Y_1(\alpha_{0,1})}{J_1(\alpha_{0,1})} \approx - \frac{4}{\pi} \biggl(\frac{1}{\alpha_{0,1}^2} 
- \frac12 (\ln(\alpha_{0,1}/2) + \gamma) + \frac38\biggr) + \ldots
\end{equation}
we get in the leading order in $\rho$
\begin{equation}  \label{eq:alpha0_rho0}
\alpha_{0,1} \approx \frac{\sqrt{2}}{\sqrt{\ln(1/\rho) - 3/4}}  \quad (\rho \to 0).
\end{equation}
One sees that $\alpha_{0,1}$ indeed vanishes with $\rho$ but extremely
slowly.  This relation is also consistent with the general asymptotic
behavior of the Laplacian eigenvalues in planar domains
\cite{Mazya85,Ward93}.

Similarly, the associated eigenfunctions $v_{k,1}(\r)$ approach
$\sqrt{2}\, v_{k,2}(\r)$ as $R_1 \to 0$, but this approach can be
logarithmically slow too \cite{Ward93}.  As a consequence, the matrix
$A$, whose elements were defined in Eq. (\ref{eq:A_def}) as a weighted
scalar product of these functions, approaches $I/\sqrt{2}$, where $I$
is the identity matrix.  In the leading order, one gets thus
%
\begin{equation}
W_{k,k'} \approx \delta_{k,k'} B_k^{(1)} B_{k'}^{(2)} \,,
\end{equation}
and the diagonal structure of this matrix allows for the explicit
inversion of $I+W$.  We get therefore
\begin{equation}
c_{0,2} \approx \frac{1}{1 + \frac{R_2}{L_2} B_0^{(1)}}  \qquad (\rho \to 0),
\end{equation}
from which Eq. (\ref{eq:c02_bis}) implies
\begin{equation} 
\zeta_\kappa(\rho,h) \approx B_0^{(1)}  \qquad (\rho \to 0).
\end{equation}
Using the definition (\ref{eq:Bn1}) of $B_n^{(1)}$, we deduce the
asymptotic approximation (\ref{eq:zeta_rho0}).  Together with
Eq. (\ref{eq:alpha0_rho0}) for $\alpha_{0,1}$, we got the fully
explicit approximation for the dimensionless offset
$\zeta_\kappa(\rho,h)$ in the limit $\rho \to 0$.

\subsection{The limit of an absorbing disk}

As $L_1 \to 0$, the pillar shrinks to an absorbing disk on the
partially reactive base.  Let us assume that $\kappa > 0$.  In this
limit, one has $B_n^{(1)} = 1/\kappa$ so that
\begin{align*} 
W_{j,n} & = \frac{2 B_{n}^{(2)}}{\kappa} 
\int\limits_\rho^1 d\r_1 \r_1 v_{j,2}(\r_1) \int\limits_\rho^1 d\r_2 v_{n,2}(\r_2) \\
& \times \underbrace{\sum\limits_{n'=0}^\infty \r_2 v_{n',1}(\r_1) v_{n',1}(\r_2)}_{=\delta(\r_1-\r_2)} \\
& = \frac{2 B_{n}^{(2)}}{\kappa}  \int\limits_\rho^1 d\r \,\r \, v_{j,2}(\r) \, v_{n,2}(\r),
\end{align*}
where we used the completeness relation for eigenfunctions
$\{v_{n,1}(\r)\}$.  The last integral can be computed exactly, in
analogy to Eq. (\ref{eq:A}):
\begin{equation}   \label{eq:Wkn_disk}
W_{j,n} = \frac{2 B_{n}^{(2)}}{\kappa}   \rho \frac{v_{j,2}(\rho) v'_{n,2}(\rho) - v_{n,2}(\rho) v'_{j,2}(\rho)}{\alpha_{n,2}^2 - \alpha_{j,2}^2}
\end{equation}
for $j\ne n$.  In turn, for $j = n$, one can use the following
identity
\begin{align*}
\int\limits dr \, r\, J_\nu^2(\alpha r) & = \frac{r^2}{2}\biggl(J_\nu^2(\alpha r) - J_{\nu-1}(\alpha r) J_{\nu+1}(\alpha r)\biggr) \\
& = \frac{r^2}{2}\biggl(J_\nu^2(\alpha r) - \biggl(\frac{\nu J_{\nu}(\alpha r)}{\alpha r}\biggr)^2 + [J'_{\nu}(\alpha r)]^2\biggr). 
\end{align*}
Setting $\nu = 0$, we find
\begin{align*}
\int\limits_{\rho}^1 d\r \, \r\, v_{n,2}^2(\r) & = \frac{r^2}{2}\biggl(v_{n,2}^2(\r) + \frac{[v'_{n,2}(\r)]^2}{\alpha_{n,2}^2}\biggr)_{\r=\rho}^1 \\
& = \frac{1}{2} - \frac{\rho^2}{2} \biggl(v_{n,2}^2(\rho) + \frac{[v'_{n,2}(\rho)]^2}{\alpha_{n,2}^2}\biggr),
\end{align*}
where we used boundary conditions.  We get then
\begin{equation}   \label{eq:Wkk_disk}
W_{n,n} = \frac{B_{n}^{(2)}}{\kappa}  \biggl(1 - \rho^2 \biggl(v_{n,2}^2(\rho) + \frac{[v'_{n,2}(\rho)]^2}{\alpha_{n,2}^2}\biggr)\biggr).
\end{equation}
We stress that the elements of the matrix $W$ are fully explicit in
this case.  The inversion of the matrix $I+W$ determines the
coefficients $\{c_{k,2}\}$ and thus the whole solution for an
absorbing disk on a partially reactive base.

In the trivial limit $\kappa = \infty$, one gets $W = 0$, from which
$c_{k,2} = \delta_{k,0}$ and one retrieves the expected solution
$u(r,z) = z/L_2$ that corresponds to the whole absorbing surface at
the bottom.  In turn, the opposite limit $\kappa\to 0$ is
mathematically more difficult as the contribution of the matrix $W$
becomes dominant but the identity matrix $I$ cannot be neglected in
$(I+W)^{-1}$ (as $W$ is expected to be non-invertible).

For the reflecting base ($\kappa = 0$), Fig. \ref{fig:J_h0}
illustrated that the offset can be very accurately approximated by the
Fock's function, see Eq. (\ref{eq:Fock}).  We expect that this result
can be deduced from our exact solution.  On the one hand, one can
attempt to undertake the limit $\kappa \to 0$ of the matrix $(I +
W')^{-1}$ by using the exact explicit form (\ref{eq:Wkn_disk},
\ref{eq:Wkk_disk}) of the matrix elements $W_{k,n}$.  On the other
hand, one can first set $\kappa = 0$ in Eq. (\ref{eq:Bn1}) to get
$B_n^{(1)} = \ctanh(\alpha_{n,1}h)/\alpha_{n,1}$ and then analyze the
asymptotic behavior $h\to 0$.  The major technical difficulty is that
one cannot replace this expression by its leading-order $B_n^{(1)}
\approx 1/(h \alpha_{n,1}^2)$.  In fact, for any {\it fixed} $h$, this
approximation is valid for $n \lesssim n_{\rm max}$ with some $n_{\rm
max}$ but one still has $B_n^{(1)} \approx 1/\alpha_{n,1}$ for $n \gg
n_{\rm max}$.  As a consequence, the limit $h\to 0$ resembles the
semi-classical asymptotic analysis of a hamiltonian $h^2 \Delta + V$
with a bounded potential $V$ in quantum mechanics, and thus requires
finer asymptotic tools.  Moreover, one generally needs very large
truncation orders to get accurate numerical results.

The same difficulty emerges in the limit $R_2\to \infty$, which is
equivalent to the double limit $\rho\to 0$ and $h\to 0$ with $h/\rho$
being fixed.

\subsection{Extensions} 
\label{sec:extensions}

While we focused on solving the particular boundary value problem
(\ref{eq:problem_u}), the proposed method can be easily extended in
several directions.

i) The Laplace equation $\Delta u = 0$ can be replaced by the modified
Helmholtz equation, $(\Delta - p/D) u = 0$.  In this case, it is
sufficient to replace $\alpha_{n,1}$ by $\alpha'_{n,1} =
(\alpha_{n,1}^2 + p/D)^{1/2}$ in Eqs. (\ref{eq:s1}, \ref{eq:Bn1}) and
$\alpha_{n,2}$ by $\alpha'_{n,2} = (\alpha_{n,2}^2 + p/D)^{1/2}$ in
Eqs. (\ref{eq:s2}, \ref{eq:Bn2}).  These replacements ensure that the
series representations (\ref{eq:u1}, \ref{eq:u2}) satisfy the modified
Helmholtz equation.  The rest of computations remains unchanged (see
details in \cite{Grebenkov_2023}).  Moreover, an inverse Laplace
transform of $u$ with respect to $p$ gives the time evolution of the
concentration governed by the diffusion equation,
\begin{equation}
\partial_t u = D \Delta u \quad \textrm{in} ~\Omega ,
\end{equation}
with the initial condition $u(r,z,t=0) = 0$, and the same boundary
conditions as in Eqs. (\ref{eq:problem_u}).  This equation also
describes heat propagation from the ``hot'' top plane at $z = L_2$
(kept at a constant temperature $1$) to the bottom spiky support (kept
at a constant temperature $0$).

ii) The homogeneous boundary condition $u(r,L_2) = 1$ on the top disk
can be replaced by $u(r,L_2) = f(r)$ with a given function $f(r)$.
For this purpose, it is sufficient to replace the right-hand side of
Eq. (\ref{eq:system}) by
\begin{equation}
\int\limits_0^1 d\r \, \r \, v_{n,2}(\r) \, f(\r R_2) .
\end{equation}
One easily obtains therefore a general form of the source term.  

Moreover, setting $f(r) = 0$ yields a homogeneous system of linear
equations, which has either none or infinitely many nontrivial
solutions, according to whether $\det(I + W) \ne 0$ or not.  For the
Laplace equation, the only solution is $c_{k,2} = 0$ for all $k$,
i.e., $u(r,z) \equiv 0$.  In turn, for the modified Helmholtz
equation, the condition $\det(I+W) = 0$ allows one to determine the
eigenvalues of the Laplace operator in this domain; the associated
eigenfunctions are given $u(r,z)$.

iii) The Dirichlet boundary condition $u(r,L_2) = 1$ can be replaced
by Neumann boundary condition $(\partial_z u(r,z))_{z=L_2} = 0$.  For
this purpose, the functions $s_{n,2}(z)$ from Eq. (\ref{eq:s2}) should
be replaced by
\begin{equation}
s_{n,2}^{\rm N}(z) = \frac{\cosh(\alpha_{n,2}(L_2-z)/R_2)}{\cosh(\alpha_{n,2} L_2/R_2)} \,.
\end{equation}
Accordingly, one would have
\begin{equation}
B_n^{(2),\rm N} = - R_2 \bigl(\partial_z s_{n,2}^{{\rm N}}(z)\bigr)_{z=0} = \alpha_{n,2} \tanh(\alpha_{n,2} L_2/R_2).
\end{equation}
In addition, one removes $1$ from Eq. (\ref{eq:u2}) and sets the
right-hand side of Eq. (\ref{eq:system}) to $0$.  One can similarly
treat the Robin boundary condition at $z = L_2$.  Together with the
above modification for the modified Helmholtz equation, the solution
$u(r,z)$ of the modified problem can be interpreted as the Laplace
transform of the probability density function (PDF) of the
first-passage time to the absorbing pillar.

(iv) The mode matching method is also applicable to more complicated
geometric settings.  In particular, one can consider a cylindrical
pillar of arbitrary cross section $\Gamma_1$ inside a collinear outer
tube of arbitrary cross section $\Gamma_2$ (such that $\Gamma_1
\subset \Gamma_2$).  In this case, the radial functions $v_{n,1}(r)$
are replaced by the normalized eigenfunctions of the two-dimensional
Laplace operator in the annular domain $\Gamma_2 \backslash \Gamma_1$,
with Dirichlet boundary condition on $\partial \Gamma_1$ and Neumann
boundary condition on $\partial \Gamma_2$.  The associated eigenvalues
$\lambda_{n,1}$ give $\alpha_{n,1}/R_2 =
\sqrt{\lambda_{n,1}}$.  Similarly, $v_{n,2}(r)$ are replaced by the
normalized eigenfunctions of the two-dimensional Laplace operator in
$\Gamma_2$, with Neumann boundary condition on $\partial
\Gamma_2$, and $\alpha_{n,2}/R_2 = \sqrt{\lambda_{n,2}}$.  The rest of
computations remains unchanged; in particular, the total flux is still
given by Eq. (\ref{eq:Jexact}), in which $\pi R_2^2$ is replaced by
the area $|\Gamma_2|$ of the tube cross section, and the offset
parameter $z_0$ is still given by Eq. (\ref{eq:z0}).  Even though this
extension is elementary, the solution looses its analytic character
because the eigenfunctions and eigenvalues of the Laplace operator are
generally not known explicitly \cite{Grebenkov13}, except for a few
cases such as a disk or an annulus between two concentric circles (as
we considered in this paper).  Nevertheless, the universal structure
of this extended solution allows one to draw some general conclusions,
such as the form (\ref{eq:Jexact}) of the total flux.  Moreover, the
average of the solution over the cross section is still linear, as in
Eq. (\ref{eq:u2_bar}).  This is a consequence of the fact that the
ground eigenfunction $v_{0,2}$ is constant, while the corresponding
eigenvalue is $\lambda_{0,2} = 0$, whatever the cross section of the
tube.

The last observation provides a simple way to compute the total flux
by Monte Carlo simulations.  According to Eq. (\ref{eq:u2_bar}), one
has $\overline{u}_2(0) = 1 - c_{0,2}$ so that the total flux is
\begin{equation}
J = \frac{|\Gamma_2| D \c_0}{L_2} \bigl(1 - \overline{u}_2(0)\bigr).
\end{equation}
Let us recall that the solution $u(\x)$ can be interpreted as the
splitting probability: if a particle starts from $\x$, this is the
probability of hitting the top cross-section at $z = L_2$ before being
absorbed on the pillar (or on the base).  If the starting point is
uniformly distributed on the tube cross section at $z = 0$, the
average $\overline{u}_2(0)$ is again the splitting probability, which
can be easily accessed via Monte Carlo simulations.  For this purpose,
one can generate $M$ random trajectories of a diffusing particle
started uniformly at $z = 0$, and count how many of them, $M_{\rm
top}$, reached the level $z = L_2$ before being absorbed.  When $M$ is
large, $M_{\rm top}/M$ approximates $\overline{u}_2(0)$, and thus
determines the total flux.

\end{document}